\documentclass[11pt]{article}
\usepackage[utf8]{inputenc}
\usepackage{array}
\usepackage[T1]{fontenc}
\usepackage{tabu}
\usepackage{mathtools}
\usepackage{bm}
\usepackage{authblk}
\usepackage{amsfonts}
\usepackage{color}
\usepackage{tabu}
\usepackage{colortbl}
\usepackage{hyperref}

\usepackage{algpseudocode}
\usepackage{subcaption}
\usepackage{algorithm}

\title{Greedy routing optimisation in hyperbolic networks}

\author[1]{Bendegúz Sulyok}
\author[1,2,*]{Gergely Palla}

\affil[1]{Dept. of Biological Physics, Eötvös Loránd University, H-1117 Budapest, Pázmány P. stny. 1/A, Hungary}
\affil[2]{Data-Driven Health Division of National Laboratory for Health Security, Health Services Management Training Centre, Semmelweis University,  H-1125, Kútvölgyi út 2, Budapest, Hungary}

\affil[*]{pallag@hal.elte.hu}

\begin{document}
\maketitle

\begin{abstract}
Finding the optimal embedding of networks into low-dimensional hyperbolic spaces is a challenge that received considerable interest in recent years, with several different approaches proposed in the literature. In general, these methods take advantage of the exponentially growing volume of the hyperbolic space as a function of the radius from the origin, allowing a (roughly) uniform spatial distribution of the nodes even for scale-free small-world networks, where the connection probability between pairs decays with hyperbolic distance. One of the motivations behind hyperbolic embedding is that optimal placement of the nodes in a hyperbolic space is widely thought to enable efficient navigation on top of the network. According to that, one of the measures that can be used to quantify the quality of different embeddings is given by the fraction of successful greedy paths following a simple navigation protocol based on the hyperbolic coordinates. In the present work, we develop an optimisation scheme for this score in the native disk representation of the hyperbolic space. This optimisation algorithm can be either used as an embedding method alone, or it can be applied to improve this score for embeddings obtained from other methods. According to our tests on synthetic and real networks, the proposed optimisation can considerably enhance the success rate of greedy paths in several cases, improving the given embedding from the point of view of navigability. 
\end{abstract}

\flushbottom
\maketitle
%
%
\thispagestyle{empty}


\section*{Introduction}
Network theory has become ubiquitous in the analysis of complex systems ranging from molecular interactions up to the level of the global economy or the entire society \cite{Laci_revmod,Dorog_book,Newman_Barabasi_Watts}. One of the very notable approaches in modelling the statistical features of the web of connections between parts of a complex system is given by hyperbolic networks\cite{hyperGeomBasics,PSO,EPSO_HyperMap,S1,S1H2_Mercator,our_embedding,Boguna_network_geom_review}, where nodes are placed in a hyperbolic space and are connected according to a probability that is decreasing with the hyperbolic distance. These models can generate scale-free, highly clustered and small-world random graphs, reproducing the most important universal features of networks representing complex systems. In addition, the often observed modular structure of real-world networks\cite{Fortunato_coms,Fortunato_Hric_coms,Cherifi_coms} can also be easily grasped by these approaches
\cite{GPA_PSOsoftComms,nPSO,our_hyp_coms,our_modular_pso}. 

Probably the most well-known hyperbolic model is the popularity-similarity optimisation (PSO) model\cite{PSO}, where the nodes are introduced one by one at logarithmically increasing radial coordinates and random uniform angular coordinates in the native disk representation of the 2d hyperbolic space, and the connection probability between nodes is decaying according to a Fermi-function, depending on the hyperbolic distance and a temperature-like parameter. Another very notable approach is provided by the random hyperbolic graph\cite{hyperGeomBasics}, where the network is static and is obtained by placing the nodes at random onto the native disk and connecting them according to a connection probability that is decaying with the hyperbolic distance in a similar fashion as mentioned above. Several variations and generalisations of these seminal models were proposed over the years by e.g., adding slight modifications to the linking procedures\cite{EPSO_HyperMap,S1,S1H2_Mercator,our_embedding}, extending the approaches to higher dimensions\cite{firstDdimHypModel,RHG_d_dim_mathematics,RHG_d_dim_krioukov,our_d_dim_PSO}, or incorporating tunable community structures\cite{nPSO,GPA_PSOsoftComms}. 

The success of hyperbolic models inherently brought with itself the interest towards the inverse problem as well, where the task is to find an optimal arrangement of the nodes in the hyperbolic space based on a given input network data. The first ideas about the hyperbolic embedding of networks appeared in Ref.~\cite{Boguna_Krioukov_Internet_2010}, which was followed by the development of various different approaches later. A very natural idea that comes up is likelihood optimisation with respect to a hyperbolic model, as implemented in Hypermap~\cite{EPSO_HyperMap}, an early method for minimising a logarithmic loss function based on the assumption that the input network was generated according to an extended version of the PSO-model. Another popular option is to apply dimension reduction techniques on matrices representing the distance relations between the nodes, leading to model-independent embeddings such as the Laplacian eigenmaps approach~\cite{Alanis-Lobato_LE_embedding}, the family of coalescent embeddings~\cite{coalescentEmbedding}, and the Hydra method~\cite{Hydra}. Dimension reduction and optimisation techniques can be also combined as proposed in the case of the Mercator method~\cite{Mercator} or when fusing the Laplacian embedding approach with E-PSO model-based optimisation~\cite{Alanis-Lobat_liekly_LE_emb}, 
or when applying a local likelihood optimisation to the output of a coalescent embedding algorithm \cite{our_embedding}. The recent generalisations of hyperbolic embedding approaches include methods for dealing with directed networks~\cite{our_dir_embedding} and the embedding of multiplex networks as well~\cite{Kleinberg_multilayer_embedding}.  

The quality of an embedding can be quantified according to several different measures, e.g., in the case of likelihood optimisation, the lowest achieved value of the loss function is a straightforward simple quality indicator. Another approach for measuring the quality of hyperbolic embeddings is focusing on greedy routing, motivated by that under optimal circumstances the links in a hyperbolic network tend to follow the geodesic lines~\cite{hyperGeomBasics}, enabling an efficient routing based on the hyperbolic coordinates~\cite{Boguna_2009_nat_phys}. The idea of greedy routing on networks embedded in a geometric space in general goes back to the pioneering work by Kleinberg~\cite{Kleinberg_greedy_routing}, considering a navigation protocol where we always proceed to the neighbour that is the closest to the destination node according to the distance defined in the given geometric space. Naturally, for networks embedded in a hyperbolic space, this greedy routing path is based on the hyperbolic distance between the nodes~\cite{Boguna_2009_nat_phys}. The routing stops either when the target has been reached, or when hoping onto an already visited node, meaning that instead of reaching the target the path ends in a cycle and the greedy routing is unsuccessful~\cite{Boguna_2009_nat_phys,Boguna_Krioukov_Internet_2010}. 

Although hyperbolic networks are usually considered to be very suitable for greedy routing~\cite{hyperGeomBasics,EPSO_HyperMap,Boguna_2009_nat_phys,Boguna_Krioukov_Internet_2010,coalescentEmbedding}, still, in most cases, the greedy paths are not 100\% successful in reaching the target, as pointed out by a recent study also raising some concerns regarding the widely believed high congruence between hyperbolic networks and their underlying space\cite{Carlo_Nat_coms_hyp_congruency}. Motivated by that, here we develop an optimisation procedure for increasing the efficiency of greedy routing on the native disk representation of the hyperbolic space, and apply it to both PSO networks generated in the hyperbolic space itself and real networks embedded in the native disk. Besides the study of the achievable increase in the greedy routing efficiency we also examine how the optimisation affects further properties of the embedding. 

\section*{Results}
\subsection*{Preliminaries}

The efficiency of greedy routing is affected by both the fraction of successful greedy paths and the length of these successful paths. A quantity for quantifying the greedy routing efficiency that grasps both of these factors is given by the greedy routing score~\cite{coalescentEmbedding}
\begin{equation}
{\rm GR}(\{r_i,\theta_i\})= \frac{1}{N(N-1)}\cdot\sum\limits_{s\in N}\,\,\sum\limits_{t\in N, t\neq s}\frac{\ell_{s\rightarrow t}^{\mathrm{(SP)}}}{\ell_{s\rightarrow t}^{\mathrm{(GR)}}}, 
\label{eq:GRscoreDef}
\end{equation}
where $(\{r_i,\theta_i\})$ refers to the actual node positions given with the help of radial and angular coordinates, $N$ is the number of nodes and the summations take into account all possible source-target pairs with $\ell_{s\rightarrow t}^{\mathrm{(SP)}}$ and $\ell_{s\rightarrow t}^{\mathrm{(GR)}}$ denoting the length of shortest and greedy paths respectively (where $\ell_{s\rightarrow t}^{\mathrm{(GR)}}=\infty$ if the routing is unsuccessful). (The details of the calculation of the above measure are described in detail in the Methods).

Unsuccessful paths, where the routing protocol enters a cycle instead of reaching the target usually cause a more serious problem compared to the issues raised by the fact that the length of the successful paths can be sub-optimal. Hence, the main priority of our optimisation algorithm is to increase the fraction of successful paths. A quantity focusing only on the success of the paths was introduced in the literature as the success ratio, $p_{\rm s}$, corresponding to the fraction of the successful greedy paths when all possible source and target pairs are considered\cite{Boguna_2009_nat_phys}. This can be written in a similar fashion to (\ref{eq:GRscoreDef}) in the form of
\begin{equation}
    p_{\rm s}(\{r_i, \theta_i\})
    =
    \frac{1}{N (N-1)}
    \sum_{s \in V} \sum_{\substack{t \in V \\ t \neq s}} \delta_{s \rightarrow t},
    \label{eq:GSDef}
\end{equation}
where $\delta_{s \rightarrow t} = 1$ if the greedy routing from the source \textit{s} to the target \textit{t} is successful, and otherwise $\delta_{s \rightarrow t} = 0$. 

\begin{figure}[hbt]
    \centering
    \includegraphics[width=0.75\textwidth]{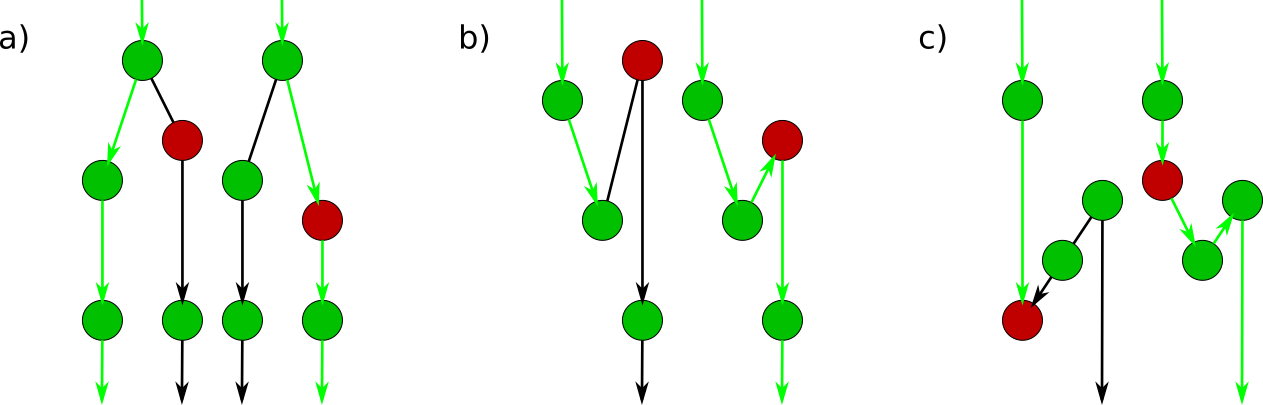}
    \caption{ {\bf Possibly elementary changes in greedy routes when displacing a single node}. In all panels, the target of the path is positioned at an arbitrarily large distance downward, and the displaced node is highlighted in blue. Panel a) shows the case of re-routing, whereas panels b) and c) display clogging and unclogging.}
    \label{fig:greedy_routing_elementary_changes}
\end{figure}
Naturally, the optimisation algorithm must displace at least a part of the nodes to induce changes in the routing of the paths. Before describing the details of the optimisation it is instructive to first consider the possible effects coming from the displacement of a single node from the point of view of successful and unsuccessful paths as listed in Fig.\ref{fig:greedy_routing_elementary_changes}. The simplest scenario, where the displacement of a given node has no effect on the considered greedy path is not shown in the figure. In case the displaced node is not part of the greedy path before the change but is adjacent to one of the nodes in the path, the displacement may reroute the path by directing it through the displaced node, as shown in Fig\ref{fig:greedy_routing_elementary_changes}a. If the displaced node is adjacent to the last node of an unsuccessful greedy path, the displacement might eliminate the clogging, as indicated in Fig.\ref{fig:greedy_routing_elementary_changes}b. The clogging in a  greedy path can be relieved also by displacing the before the last node, as shown in Fig.\ref{fig:greedy_routing_elementary_changes}c. It is important to note that for all the different changes in the routing that are listed in Fig.\ref{fig:greedy_routing_elementary_changes} the "inverse" may also happen at the displacement of nodes (e.g., instead of eliminating the clogging in an  unsuccessful path we may accidentally induce the clogging of a previously successful path). 

\begin{figure}[hbt]
    \centering
    \includegraphics[width=\textwidth]{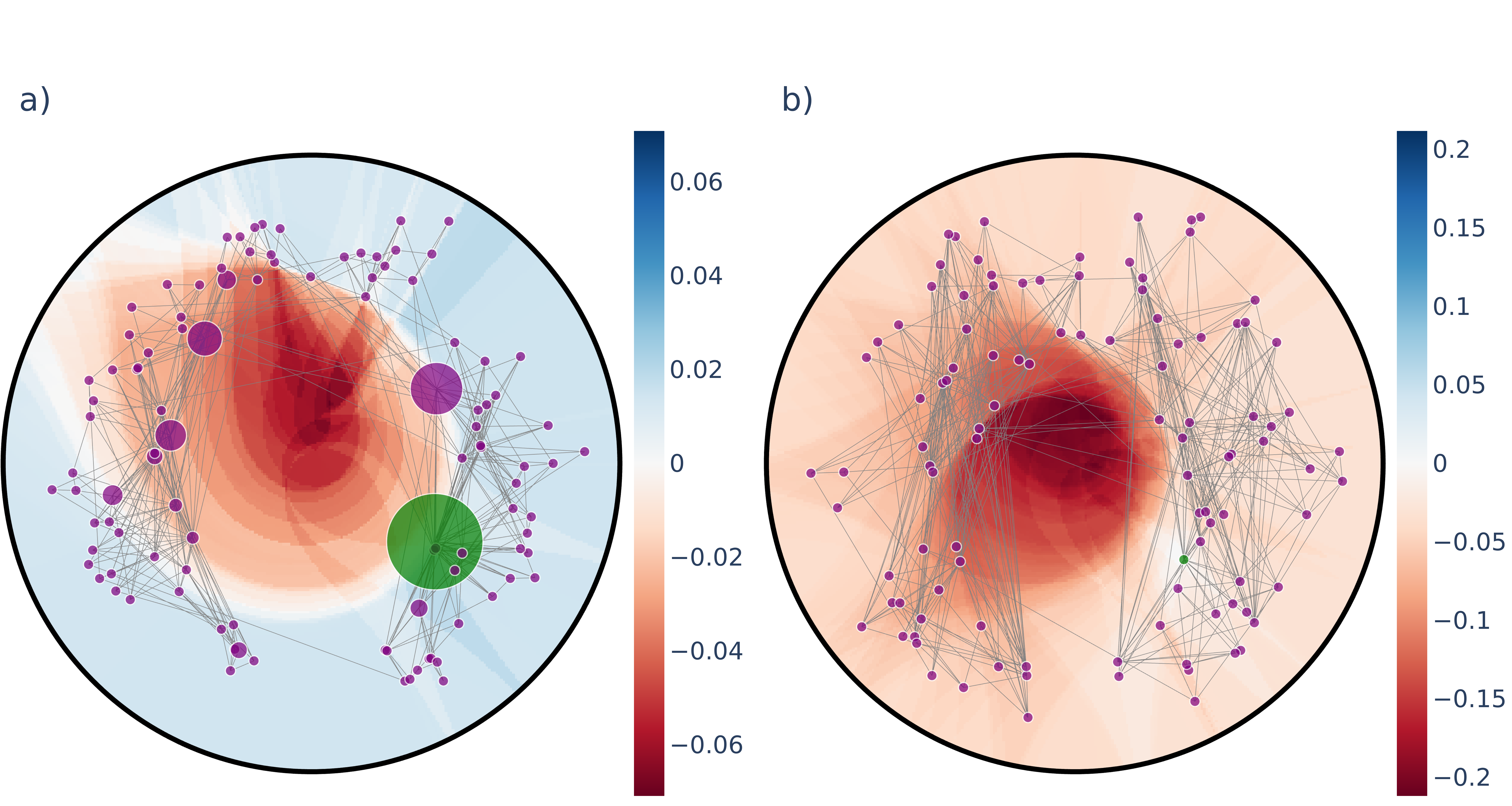}
    \caption{{\bf Change in the success ratio under relocation of a single node.} The node to be relocated is shown in green and the background colour indicates the change in $p_{\rm s}(\{r_i,\theta_i\})$ defined in Eq.(\ref{eq:GSDef}) if the node is moved to the given position. The node size is proportional to the number of unsuccessful paths ending at the given node. In panel a) we show the results for a PSO network of $N=1024$ nodes at the beginning of the optimisation process, whereas panel b) depicts the heat-map at the end of the optimisation.}
    \label{fig:GS_heatmap}
\end{figure}

To illustrate that the displacement of even a single node can change the greedy routing in a non-trivial way, in Fig.\ref{fig:GS_heatmap}. we show heat maps of the change in the success ratio when relocating a randomly chosen node within the disk defined by the outermost node in a PSO network. According to Fig.\ref{fig:GS_heatmap}a, the regions where the change is positive (indicated by the different shades of blue) form intricate patterns together with the regions where the change becomes negative (shown by the different shades of red). However, under a greedy routing optimisation process we expect that the target regions for relocation with a positive change in $p{\rm s}(\{r_i,\theta_i\})$ will shrink. This is consistent with Fig.\ref{fig:GS_heatmap}b, showing the achievable change in the success ratio for relocating the same node in the same network as in Fig.\ref{fig:GS_heatmap}a at the end of our optimisation algorithm, where basically no positive change can be obtained by relocating the chosen node. 

\subsection*{Greedy routing optimisation algorithm}

Our approach to optimising the greedy routing paths is inspired by the concept of simulated annealing~\cite{Kirkpatrick_simmulated_annealing} in statistical physics. In this general framework, the optimisation task is transformed into the problem of finding the energy minimum in a complicated energy landscape over the parameter space, and a local optimum is found with the help of a Markov-chain Monte-Carlo method. During this optimisation, random moves are considered in the parameter space, and the acceptance probability of a given move depends on the energy difference between the two settings and also on a parameter $T$ analogous to the temperature. By starting the annealing procedure at high temperatures almost all moves are accepted, allowing the exploration of the parameter space, whereas the gradual "cooling" of the system by lowering the $T$  suppresses the acceptance of moves that increase the energy and eventually drives the algorithm into a local energy minimum.

In our case, the "energy" for a given set of node coordinates $\{r_i,\theta_i\}$ is defined as 
\begin{equation}
    E(\{r_i,\theta_i\}) =1-p_{\rm s}(\{r_i,\theta_i\}),
\end{equation}
where $p_{\rm s}(\{r_i,\theta_i\})$ is calculated according to (\ref{eq:GSDef}) and the energy difference for a transition from $\{r_i,\theta_i\}$ to $\{r'_i,\theta_i'\}$ is simply $\Delta E=E(\{r'_i,\theta_i'\})-E(\{r_i,\theta_i\})=p_{\rm s}(\{r_i,\theta_i\})-p_{\rm s}(\{r'_i,\theta'_i\})$. When sampling new coordinate settings, the acceptance probability of an actual transition from the current state to the next one follows the Metropolis-Hastings rule\cite{Metropolis_orig_paper,Hastings_paper} as
\begin{equation}
    P\left(\{r_i,\theta_i\}\rightarrow \{r_i',\theta_i'\}\right)=\left\lbrace\begin{array}{ll}
        1 & \mbox{ if } \Delta E < 0,\\
        {\mathrm e}^{-\frac{\Delta E}{T}}={\mathrm e}^{\frac{p_{\rm s}(\{r'_i,\theta'_i\})-p_{\rm s}(\{r_i,\theta_i\})}{T}} &  \mbox{ otherwise. }
    \end{array} \right.
\end{equation}

The natural question arising at this point is how to sample from the possible arrangements of the nodes in the native disk. Given any current set of the node positions $\{r_i,\theta_i\}$, the simple choice we take for gaining a new sample $\{r'_i,\theta'_i\}$ is to displace a single node only (as already illustrated in Fig.\ref{fig:GS_heatmap}), where the new position for the chosen node is drawn from a uni-modal distribution centred on the current position of the node (where the details of the distribution are given in the Methods). The pseudo-code for optimising the success ratio in this approach is given in Algorithm \ref{alg:gr_anneal}.
\begin{algorithm}
  \caption{Greedy routing annealing}\label{alg:gr_anneal}
  \begin{algorithmic}[1]
    \Procedure{GRS}{$G(V, E), initial \textunderscore coord$}
    \State $coord \gets initial \textunderscore coord$
    \State $p_{\rm s} \gets compute \textunderscore greedy \textunderscore success \textunderscore ratio()$
      \For{$step \in range(0, total \textunderscore steps)$}
        \State $u \gets select \textunderscore node \textunderscore for \textunderscore move()$
        \State $new \textunderscore coord \gets update \textunderscore node \textunderscore position()$
        \State $new \textunderscore p_{\rm s} \gets compute \textunderscore greedy \textunderscore success \textunderscore ratio()$
        \State $dE \gets p_{\rm s} - new \textunderscore p_{\rm s}$
        \If{dE < 0 \textbf{or} Rand() < exp(-dE / T) }
            \State $coord \gets new \textunderscore coord$
            \State $p_{\rm s} \gets new \textunderscore p_{\rm s}$
        \EndIf
      \EndFor
      \State \textbf{return} $coord$
    \EndProcedure
  \end{algorithmic}
\end{algorithm}

The overall framework for greedy routing optimisation as defined above allows several possibilities for choosing the node to be moved, e.g., we can choose uniformly at random from all the nodes, or choose according to a probability depending on some structural property (e.g., the degree), or choose according to a probability depending on the number of clogged greedy routs containing the given node, etc.  The most straightforward approach is the uniform random choice, which does not require any additional information. In the following, we shall refer to the sampling procedure according to this choice as "random sampling". In our studies, we also tested annealing procedures where the probability to be chosen was proportional to the node degree (the corresponding sampling method shall be referred to as "degree dependent sampling"), or to the number of clogged greedy paths starting from the given node ("clogged source sampling") or to the number of clogged greedy paths targeting the given node ("clogged target sampling").


\subsection*{Embedding via greedy routing optimisation}

\begin{figure}[h!]
    \centering
    \includegraphics[width=0.75\textwidth]{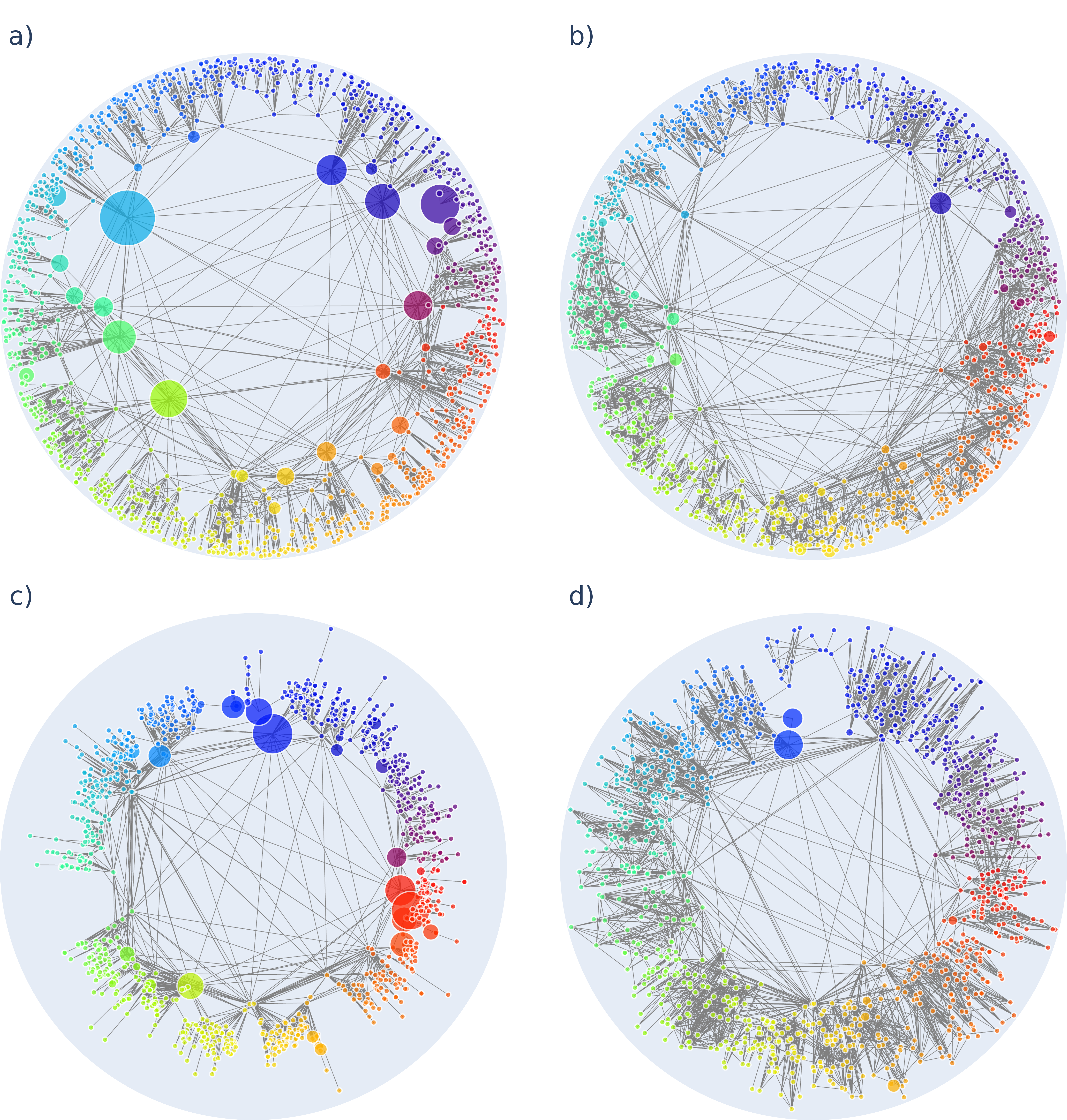}
    \caption{{\bf Improving the greedy routing in a PSO network.} The layout of the network at the end of the network generation process according to the PSO model ($N=1024$, $m=4$, $\beta=0.5$, $T=0.1$) is shown in panel (a), whereas the modified layout according to the optimisation algorithm is displayed in panel (b). In both panels the node size is proportional to the number of  greedy routes prematurely ending at a given node, whereas the node colors indicate the initial angular coordinates. In panel (c) we show the embedding of the same network according to Mercator\cite{Mercator}, where the colouring of the nodes is adapted to this new starting layout. Panel (d) depicts the result obtained by optimising the layout shown in panel (c).}
    \label{fig:layout_comparison}
\end{figure}

The greedy routing optimisation procedure we described can be viewed both as a hyperbolic embedding algorithm (when starting from random node positions) and also as an auxiliary method for improving embeddings obtained by other hyperbolic embedding algorithms. To illustrate the kind of results that can be expected from our annealing framework, in Fig.~\ref{fig:layout_comparison}.~ we show layouts of a network of $N=1024$ nodes generated by the PSO model both before and after the optimisation. The parameters of the PSO model were set to $m=4$ (controlling the average degree as $\left<  k\right>=2m$), $\beta=0.5$ (governing a gradual outward shift of the nodes during the network generation, leading to a tunable degree decay exponent $\gamma=1+\frac{1}{\beta}$), and $T=0.1$ (corresponding to a temperature-like parameter, controlling the clustering coefficient). In Fig.~\ref{fig:layout_comparison}a we show the original layout of the network generated by the PSO model, where the node size is proportional to the number of unsuccessful greedy paths ending at the given node and the colouring of the nodes indicates simply their angular coordinates. In comparison, Fig.~\ref{fig:layout_comparison}b displays the network after applying our optimisation framework. Apparently, the node sizes are substantially smaller compared to Fig.~\ref{fig:layout_comparison}a, thus, our optimisation is indeed doing its job by modifying the node arrangement in such a way that the number of unsuccessful greedy paths is reduced. The colouring of the nodes in Fig.~\ref{fig:layout_comparison}b is indicating their original angular coordinate as given in Fig.~\ref{fig:layout_comparison}a and according to that, the overall layout of the network remained quite similar to the original one. In panels Fig.~\ref{fig:layout_comparison}c-d we show a similar comparison between the same PSO network embedded first by the Mercator algorithm\cite{Mercator} (Fig~\ref{fig:layout_comparison}c), and then optimised according to the simulated annealing procedure proposed in the present work (Fig~\ref{fig:layout_comparison}d). Again, the node sizes are seemingly smaller in Fig~\ref{fig:layout_comparison}d compared to ~\ref{fig:layout_comparison}c, thus, our optimisation has improved the greedy navigability of the embedding. Meanwhile, based on the colouring of the nodes (indicating this time the angular coordinate in the Mercator embedding) the organisation of the network on a global scale has remained mostly intact.

\begin{figure}[h!]
    \centering
    \includegraphics[width=\textwidth]{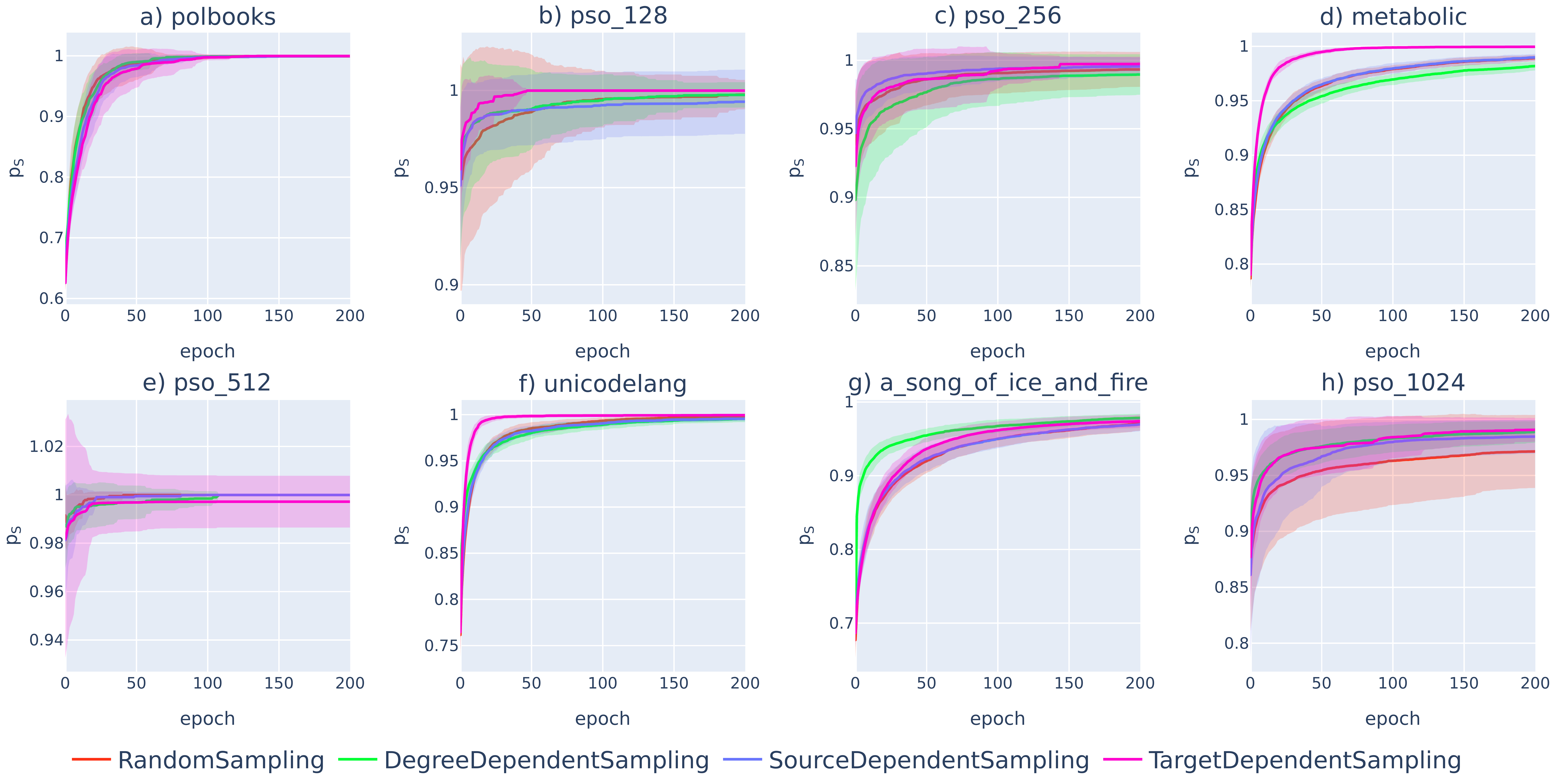}
    \caption{ {\bf Improvement of the success ratio during the optimisation.} We plot $p_{\rm s}(\{r_i,\theta_i\})$ defined in (\ref{eq:GSDef}) as a function of the number of epochs for random sampling (red), degree dependent sampling (green), clogged source dependent sampling (blue) and clogged target dependent sampling (purple). The optimisation was applied on networks initially embedded with Mercator\cite{Mercator}. The curves correspond to the average over 16 samples, and the shaded region around the curves is indicating the standard deviation. The results are shown for the network of political books ($N=105$, $L=441$) in panel (a), for a PSO network with $N=128$ nodes and $L=618$ links in panel (b), a PSO network with $N=256$ nodes and $L=1178$ links in panel (c), the metabolic network ($N=453$, $L=2025$) in panel (d), a PSO network with $N=512$ nodes and $L=2264$ links in panel (e), the unicodelang  network ($N=858$, $L=1245$) in panel (f), the network between fictional characters ($N=796$, $L=2823$) in panel (g) and a PSO network with $N=1024$ nodes and $L=4402$ links in panel (h).}
    \label{fig:GS_improve}
\end{figure}

We tested the performance of our optimisation approach on several synthetic and real networks, including PSO networks of size $N=128$, $N=256$, $N=512$ and $N=1024$, a co-purchasing network of political books\cite{polbooks} with $N=105$ nodes and $L=441$ links, a metabolic network\cite{metabolic} of $N=453$ nodes and $L=2025$ links, a bipartite language network\cite{unicodelang} containing countries and spoken languages $N=858$ nodes (10 disconnected nodes were discarded) and $L=1245$ links and the fictional social network\cite{asoiaf} of the characters appearing in the fantasy series "A Song of Ice and Fire" with $N=796$ nodes and $L=2823$ links. The real networks were obtained from \cite{konect}. In Fig.\ref{fig:GS_improve} we show the results by plotting the success ratio calculated according to (\ref{eq:GSDef}) as a function of the number of epochs (where one epoch corresponds to a number of iterations equal to the number of nodes). 

According to Fig.\ref{fig:GS_improve}., the fraction of successful greedy paths is increasing in all networks during the optimisation,  and in some cases, we can even achieve $p_{\rm s}(\{r_i,\theta_i\})=1$, meaning that all greedy paths become successful due to our optimisation. Naturally, both the  $p_{\rm s}(\{r_i,\theta_i\})$ at the end of the optimisation and the relative performance of the different annealing schemes vary over the studied networks. By comparing the results for different systems, it seems that the annealing scheme using target dependent sampling clearly outperformed the other annealing schemes in case of the metabolic network and the network between fictional characters. 
In contrast, for the network between fictional characters the annealing scheme relying on degree based sampling turned out to be the fastest in finding an optimal arrangement according to the success ratio.

\begin{figure}[hbt]
    \centering
    \includegraphics[width=\textwidth]{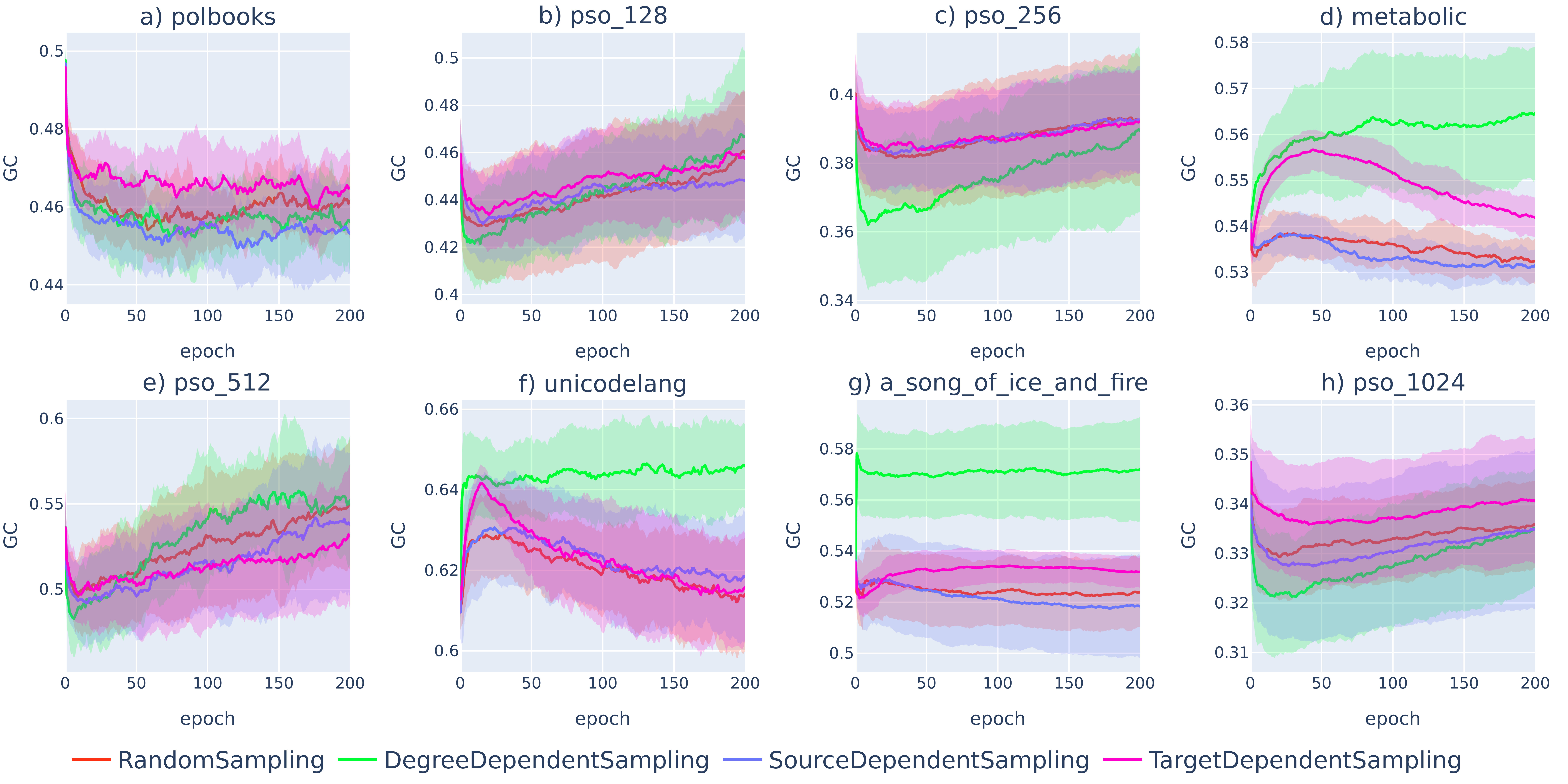}
    \caption{ {\bf Change in the geometrical congruence under the optimisation.} We show the GC defined in (\ref{eq:GC_def}) as a function of the number of epochs for the same networks (initially embedded with Mercator\cite{Mercator}) as in Fig.\ref{fig:GS_improve}, where the colour of the curves encode the annealing scheme, the value of the curve shows the average over 16 samples, and the shaded region around the curves is indicating the standard deviation. The results are shown for the network of political books in panel (a), for a PSO network with $N=128$ nodes in panel (b), a PSO network with $N=256$ in panel (c), the metabolic network in panel (d), a PSO network with $N=512$ nodes in panel (e), the unicodelang  network in panel (f), the network between fictional characters in panel (g) and a PSO network with $N=1024$ nodes in panel (h).}
    \label{fig:greedy_efficiency}
\end{figure}

Since our optimisation scheme is specifically tailored for optimising the success ratio, it is natural to ask what happens to other alternative quality measures during the simulated annealing. In Fig.~\ref{fig:greedy_efficiency}.~ we show the geometrical congruence\cite{Carlo_Nat_coms_hyp_congruency}, measuring the similarity between the geodesic distances and the projected topological distances governed by the networks structure, where the precise definition of this score is given by Eq.~(\ref{eq:GC_def}) in the Methods. 
The results show a mixed picture, as in some cases the GC-score is slightly smaller after the optimisation, whereas in other cases the annealing terminates with a higher GC-score value compared to the initial embedding. Furthermore, the curves can show a non-monotonous behaviour, having either a minimum or a maximum in the studied interval. Nevertheless, we would like to point out that the overall change in the GC-score is minor in most of the cases.

We provide more results on the proposed optimisation framework in the Supplementary Information. On the one hand, in Sect.~S1.~ we examine the behaviour of both the success ratio and the GC-score when the simulated annealing is started from a random embedding for the same networks that are studied here. According to the results, full navigability (where the success ratio is reaching $p_{\rm s}=1$) can be achieved in this case as well for some of the studied networks, however, in most cases the success ratios at the end of the optimisation are slightly below the values we observed when starting the annealing from embeddings obtained with Mercator\cite{Mercator}. In parallel, according to Fig.~S2.~,  major improvement can be observed in the geometrical congruence as well for most of the networks. 

On the other hand, in Sect.~S2.~ in the Supplementary Information we study the effect of our annealing framework on further quality scores, such as the mapping accuracy\cite{mappingAccuracyAsSPLcorr}, the area under the receiver operating characteristic curve and the area under the precision-recall curve in graph reconstruction\cite{AUROCmeaningInLinkPred,descriptionOfMeasuresOfGraRecAndLinkPred}, the greedy routing score\cite{coalescentEmbedding} as defined in Eq.(\ref{eq:GRscoreDef}) and the greedy routing efficiency\cite{Carlo_Nat_coms_hyp_congruency}. According to the results, the mapping accuracy and the quality scores related to graph reconstruction decrease when starting the optimisation from embeddings obtained with Mercator\cite{Mercator}. This can be viewed as a sort of "cost" we have to pay for achieving the improvement in the success ratio shown in Fig.~\ref{fig:GS_improve}. In parallel, the scores related to greedy routing either remained close to their original value, occasionally with a slight decay, or improved when starting the simulated annealing from an embedding obtained with Mercator\cite{Mercator}. Finally, all scores showed (in most cases major) improvement when starting the optimisation from a random embedding. Hence, the embeddings our annealing framework finds are advantageous from the point of view of a wide range of quality measures compared to random node coordinates.

\section*{Discussion}

The navigability of networks in the hyperbolic space is a topic of fundamental interest~\cite{hyperGeomBasics,Boguna_2009_nat_phys,Boguna_Krioukov_Internet_2010,Boguna_network_geom_review,Carlo_Nat_coms_hyp_congruency}. Already at the introduction of the first hyperbolic network models, one of the noted advantages of the graphs generated by these approaches was that the geodesic paths seemed to be aligned with the topological shortest paths, enabling an efficient greedy routing protocol for navigation in the network~\cite{hyperGeomBasics,Boguna_network_geom_review}. Nevertheless, in a recent work some concerns were raised regarding the congruence between hyperbolic networks and their underlying geometry~\cite{Carlo_Nat_coms_hyp_congruency}, where it was shown that in some cases, the greedy routing may lead to unsuccessful paths between a considerable fraction of the node pairs. 

Related to the above topic, in the present paper we introduced a simulated annealing framework for improving the greedy navigability of networks embedded in the hyperbolic space. According to the results, our approach is able to increase the success ratio for both synthetic graphs generated by hyperbolic network models and real networks embedded into the hyperbolic space by some graph embedding technique, reaching in some examples the maximum score, corresponding to layouts where all greedy paths are successful in reaching their targets. Although the algorithm can be used as an embedding method on its own, in practice it is more convenient to use it as an auxiliary procedure that can improve the output of other hyperbolic embedding methods due to the well-known high computation cost of simulated annealing methods. 

Besides the success ratio, we have also monitored the change of several other quality measures, including the geometrical congruence\cite{Carlo_Nat_coms_hyp_congruency}, the mapping accuracy\cite{mappingAccuracyAsSPLcorr}, the area under the receiver operating characteristic curve and the area under the precision-recall curve in graph reconstruction\cite{AUROCmeaningInLinkPred,descriptionOfMeasuresOfGraRecAndLinkPred}, the greedy routing score\cite{coalescentEmbedding} and the greedy routing efficiency\cite{Carlo_Nat_coms_hyp_congruency}. As expected, the scores related to greedy routing usually change in the positive direction under the optimisation even when starting the simulated annealing from a high quality embedding such as the output by Mercator\cite{Mercator}. In the meantime, the quality measures that are more independent from the success ratio such as the mapping accuracy or the scores measuring the performance in graph reconstruction can show a slightly decreasing tendency if the score of the initial embedding is high. The geometrical congruence showed a mixed behaviour, sometimes increasing, in other cases decreasing over the iterations. Nevertheless, the losses observed in the various measures during the experiments starting from embeddings obtained with Mercator were usually minor. In parallel, all studied quality scores showed an increasing tendency when starting the optimisation from a uniformly random node coordinates, usually achieving a significant improvement at the end of the process.

In conclusion, the simulated annealing framework we propose is a general approach for improving the navigability of hyperbolic networks. Although here we focused on increasing the fraction of successful paths, with minor changes made to the energy function, the method could also be used for optimising with respect to other quality scores instead. Our results showed that this approach was capable of maximising the success ratio in some examples, hence, similar optimisation algorithms aimed at improving the layout of hyperbolic networks with respect to alternative other measures have a great potential to be effective as well.

\section*{Code availability}

Our code is available at the following link: \href{https://github.com/bsulyok/gra}{https://github.com/bsulyok/gra}.

\section*{Methods}

\subsection*{Calculating the success ratio}

Since the calculation of the success ratio, $p_{\rm s}$, has to be carried out a large number of times, it is necessary to design  an efficient algorithm for doing so. Although various algorithms can be defined that differ in their details, the time complexity is roughly $\mathcal{O}(N^2 \log N + N^2 \left< k\right>)$ in all cases, where $N$ is the number of nodes and $\left< k\right>$ denotes the average degree of the network. Our solution builds on two main parts, one doing the search for the neighbour that is closest to the target and the other implementing the propagation along the greedy paths. The pseudo-code of this procedure is given in Algorithm \ref{alg:grs}.

\begin{algorithm}
  \caption{Greedy success ratio}\label{alg:grs}
  \begin{algorithmic}[1]
    \Procedure{$p_{\rm s}$}{$G(V, E), coord$}
    \State $score \gets 0$
      \For{$target \in V$}
        \State $next \textunderscore hop = []$
        \For{$source \in V$}
          \State $next \textunderscore hop[source] = \min{v \in N(source)}$
        \EndFor
        \For{$i \in range(\log_2(N))$}
          \For{$node \in V$}
              \State $next \textunderscore hop[node] = next \textunderscore hop[next \textunderscore hop[node]]$
          \EndFor
        \EndFor
        \For{$source \in V$}
          \If{$next \textunderscore hop[source] = target$}
            \State $score \gets score + 1$
          \EndIf
        \EndFor
      \EndFor
      \State \textbf{return} $score$
    \EndProcedure
  \end{algorithmic}
\end{algorithm}

\subsection*{Sampling the target position for displacement}

\begin{figure}
    \centering
    \includegraphics[width=\textwidth]{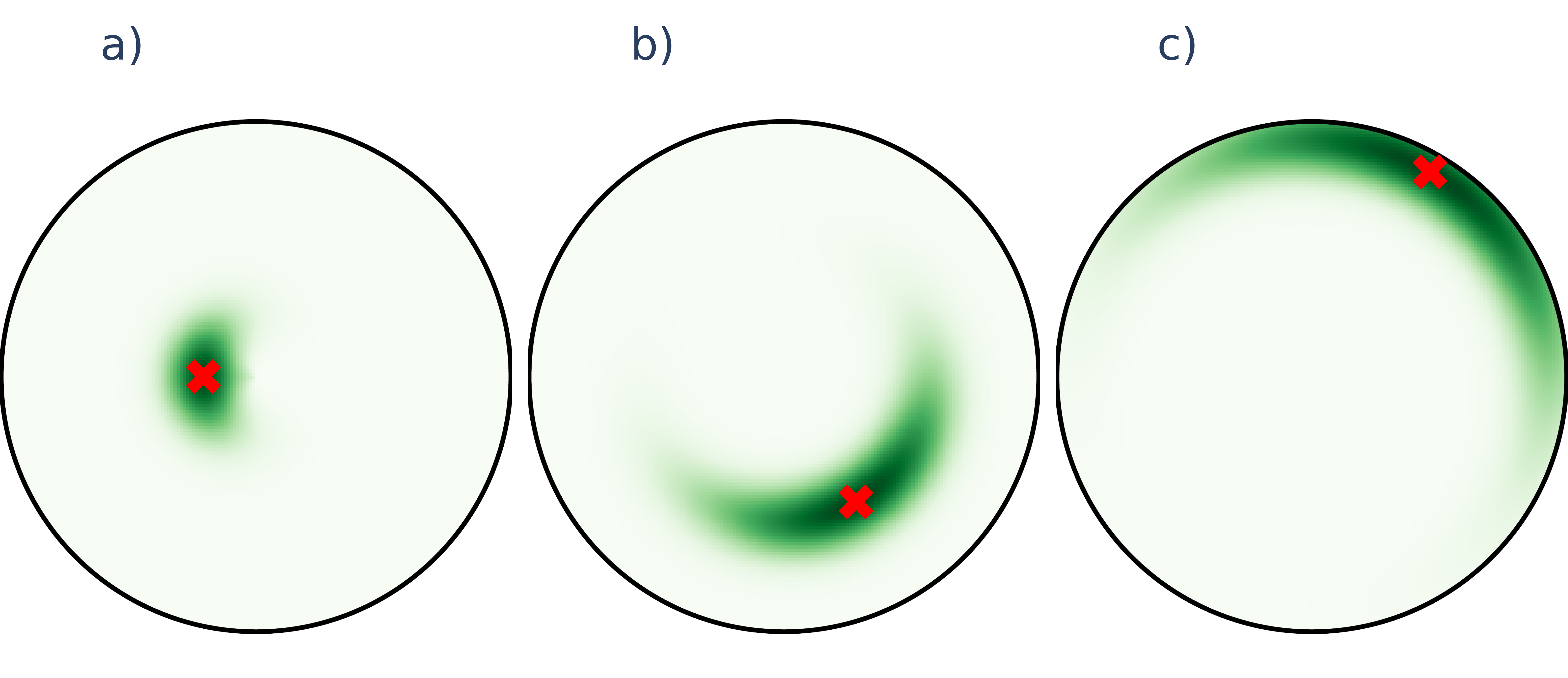}
    \caption{{ \bf Distribution of the possible new positions for a node.} The green heat map indicates the probability density and the current position of the chosen node is shown by the red marker. A network of size $N=128$ was used for generating the panels, where the chosen node is close to the disk center in panel a), at medium range from the disk center in panel b) and close to the disk periphery (far from the center) in panel c).}
    \label{fig:vertex_move_probability_heatmap}
\end{figure}

In every iteration, after selecting the node we try to displace, we also need to specify the possible new position for the selected node. This position was obtained by independently sampling a new angular coordinate from a normal distribution centered on the angular coordinate of the chosen node and a new radial coordinate from a truncated normal distribution centered on the radial coordinate of the chosen node, where the distribution was restricted to the $[0,R]$ interval with $R$ denoting the disk radius of the network. Illustrations of the resulting distribution are provided in Fig.~\ref{fig:vertex_move_probability_heatmap}. 

Although  a sampling distribution where the level lines of the density form hyperbolic circles around the chosen node might seem as a more natural choice, according to our experience the sampling distribution proposed here yields a faster increase in the success ratio over the iterations. The likely reason for this is that when the chosen node has a relatively large radial coordinate (such as e.g., in the case of Fig.~\ref{fig:vertex_move_probability_heatmap}c) our sampling distribution allows the exploration of a much larger angular region, which seems to help finding more suitable new positions.

\subsection*{The geometric congruence}
The geometric congruence was introduced in Ref.\cite{Carlo_Nat_coms_hyp_congruency} as a general measure for quantifying the alignment between the network topology and an underlying geometry. By assuming that we can evaluate the distance between any node pair, the geometric congruence is formulated as
\begin{equation}
    {\mathrm GC}(\{r_i, \theta_i\})
    =
    \frac{2}{N (N-1) - L}
    \sum_{i =1}^{N} \sum_{\substack{j=1  \\ j \notin N(i)}}^{i-1} \frac{DIST(i, j)}{PTSP(i, j)},
    \label{eq:GC_def}
\end{equation}
where $DIST(i, j)$ is the distance between node $i$ and $j$ according to the geometry, which in our case becomes the distance along the geodesic of the hyperbolic disk; and $PTSP(i, j)$ is the projected topological shortest path, that is the sum of distances along a topological shortest path (or the average of these sums if multiple topological shortest paths exist) starting on $i$ and ending on $j$. Note that the summing runs over only the  nonadjacent node pairs, since $j$ cannot be a member in the neighbour set of node $i$ (denoted by $N(i)$ in the formula).

\newpage

\begin{center}
\LARGE{\bf SUPPLEMENTARY INFORMATION}
\end{center}

\renewcommand{\thefigure}{S\arabic{figure}}
\renewcommand{\thetable}{S\arabic{table}}
\renewcommand{\theequation}{S\arabic{equation}}
\renewcommand{\thesection}{S\arabic{section}}

\setcounter{section}{0}
\setcounter{figure}{0}
\setcounter{equation}{0}
\setcounter{table}{0}

\section{Optimisation starting from random coordinates}

\begin{figure}[hbt]
    \centering
    \includegraphics[width=\textwidth]{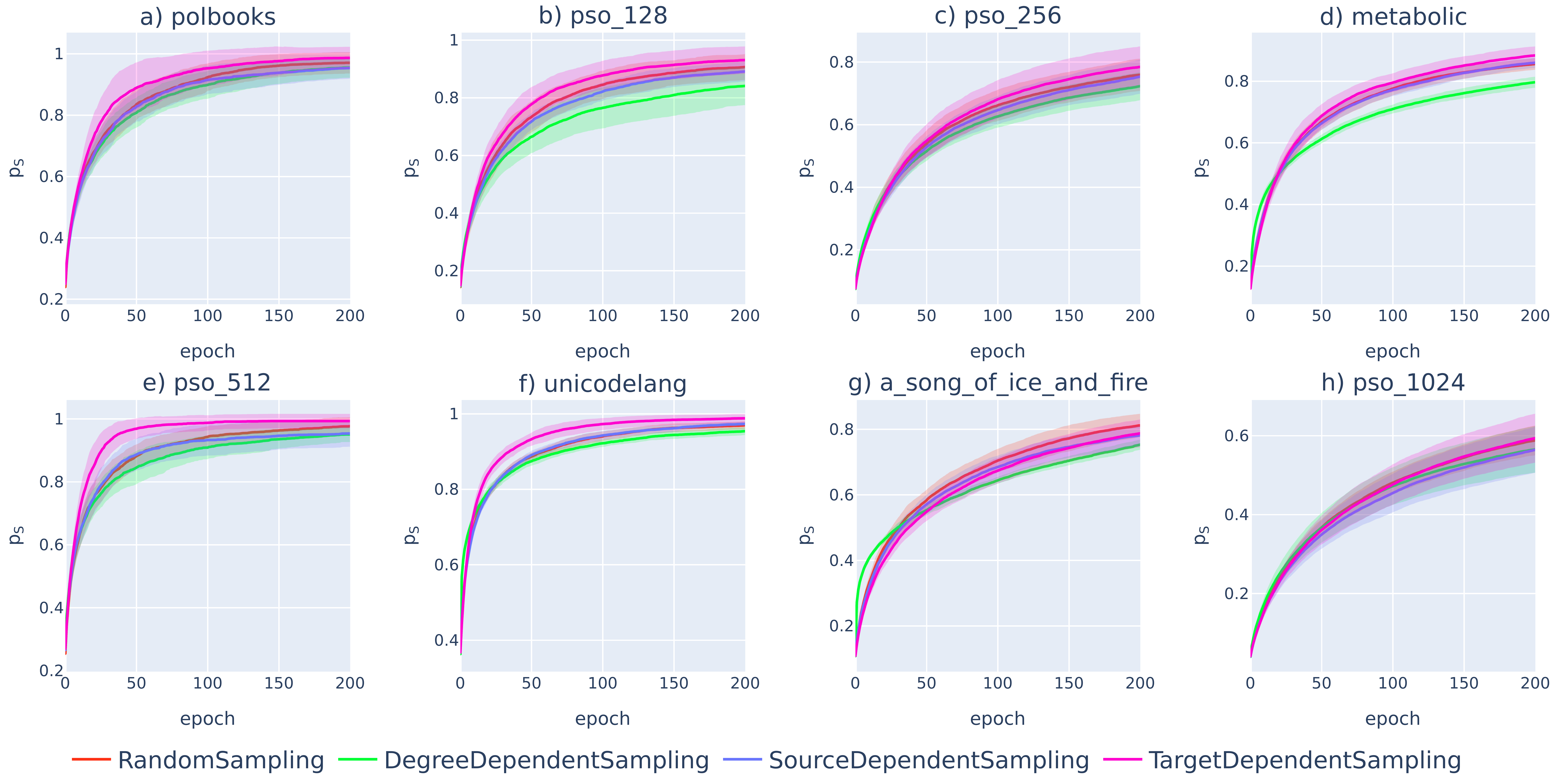}
    \caption{ {\bf Improvement of the success ratio when starting the optimisation from a random initial embedding.} We plot $p_{\rm s}(\{r_i,\theta_i\})$ defined in Eq.(2) in the main paper averaged over 16 instances as a function of the number of epochs. The colour of the curves indicates the annealing scheme and the shaded region around the curves is indicating the standard deviation. The results are shown for the network of political books in panel (a), for a PSO network with $N=128$ nodes in panel (b), a PSO network with $N=256$ nodes in panel (c), the metabolic network in panel (d), a PSO network with $N=512$ nodes in panel (e), the unicodelang  network in panel (f), the network between fictional characters in panel (g) and a PSO network with $N=1024$ nodes in panel (h).}
    \label{fig:GS_improve_random}
\end{figure}

We have also run experiments where the initial embedding of the networks was obtained by distributing the nodes uniformly at random in a native disk with radius equal to the radius of a PSO network with the same size. When applied in this manner, our method can be viewed as a full-fledged hyperbolic embedding algorithm. In Fig.~\ref{fig:GS_improve_random}.~ we show the change in the success ratio as a function of the number of epochs during the optimisation for the same networks that are studied in the main paper.
According to the results, $p_{\rm s}$ undergoes a major improvement for all studied networks, reaching even the possible maximal score for the metabolic network (Fig.~\ref{fig:GS_improve_random}d). This means that starting from a random initial state, our algorithm was able to achieve state where all greedy paths become successful. 

\begin{figure}[hbt]
    \centering
    \includegraphics[width=\textwidth]{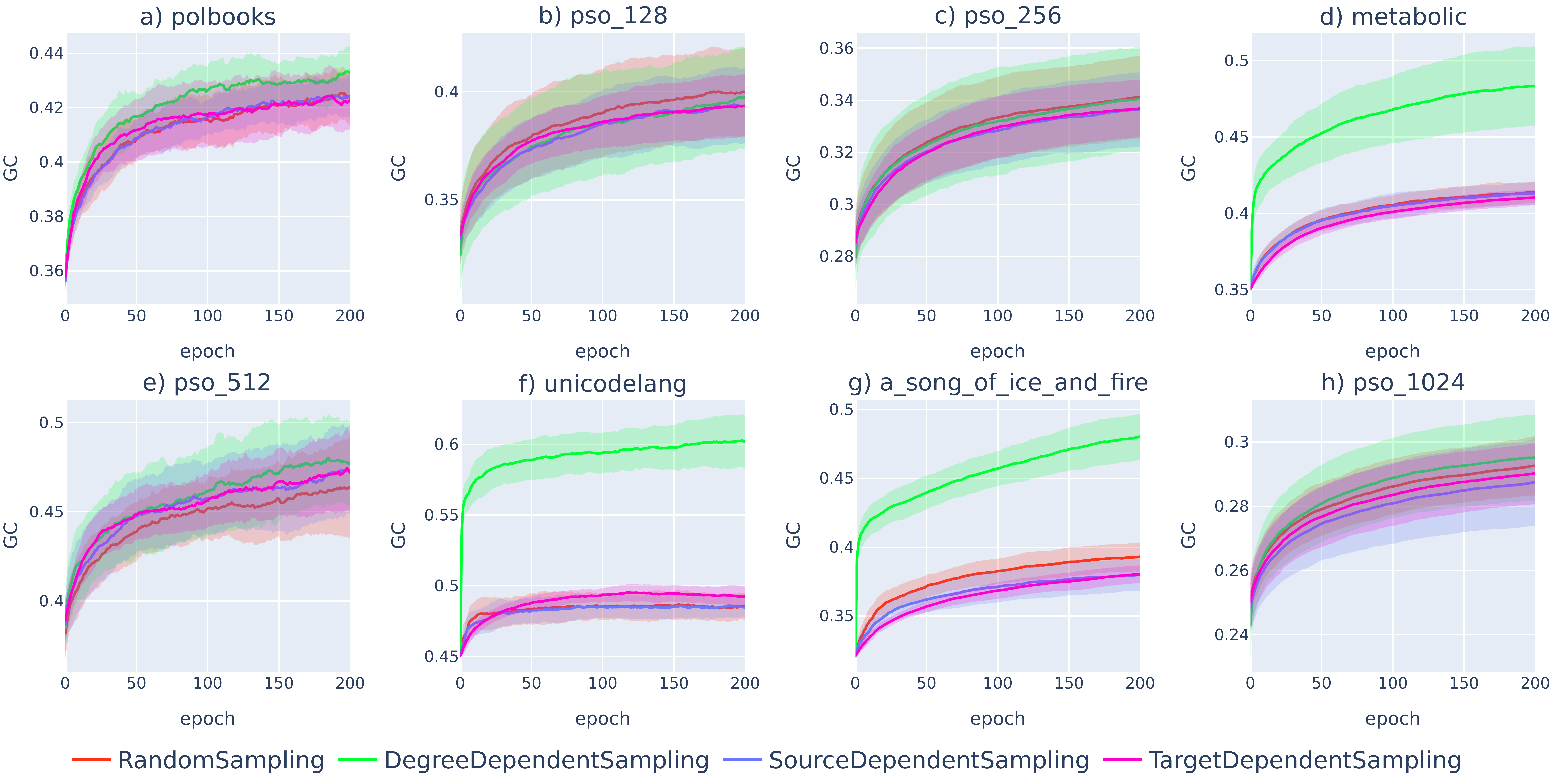}
    \caption{ {\bf Change in the geometrical congruence when the optimisation is started from a random embedding.} We plot ${\rm GC}(\{r_i,\theta_i\})$ defined in Eq.(6) in the main paper averaged over 16 samples as a function of the number of epochs. The colour of the curves indicates the annealing scheme and the shaded region around the curves is indicating the standard deviation. The results are shown for the network of political books in panel (a), for a PSO network with $N=128$ nodes in panel (b), a PSO network with $N=256$ nodes in panel (c), the metabolic network in panel (d), a PSO network with $N=512$ nodes in panel (e), the unicodelang  network in panel (f), the network between fictional characters in panel (g) and a PSO network with $N=1024$ nodes in panel (h).}
    \label{fig:GC_random}
\end{figure}

In Fig.~\ref{fig:GC_random}.~ we show the geometrical congruence as a function of the number of epochs during the same optimisation procedures. In contrast with Fig.~5.~ in the main paper, the GC-score shows an improving tendency for all networks, and the magnitude of the change is considerably larger (e.g., the largest increase was observed for the network between fictional character, where GC changed from GC=$0.321$ to GC=$0.480$ over the optimisation). Interestingly, the implementation that uses the degree-based sampling seems to strongly outperform the other alternatives for some of the networks in the study regarding this score.  

\section{Behaviour of further quality scores during the GS optimisation}

We recorded the change in several other quality scores during our numerical experiments with the proposed greedy routing optimisation procedure. In the following subsections we detail the results in terms of these measures both when starting the annealing procedure from random initial coordinates and when optimising the embedding obtained with Mercator\cite{Mercator}.

\subsection{The mapping accuracy}

The mapping accuracy is intended to measure the similarity between the geometrical distance according to the node coordinates and the shortest path length based on the network structure and is defined simply as the Spearman's correlation coefficient between these two quantities measured over all node pairs\cite{mappingAccuracyAsSPLcorr}. 

\begin{figure}[h!]
    \centering
    \includegraphics[width=\textwidth]{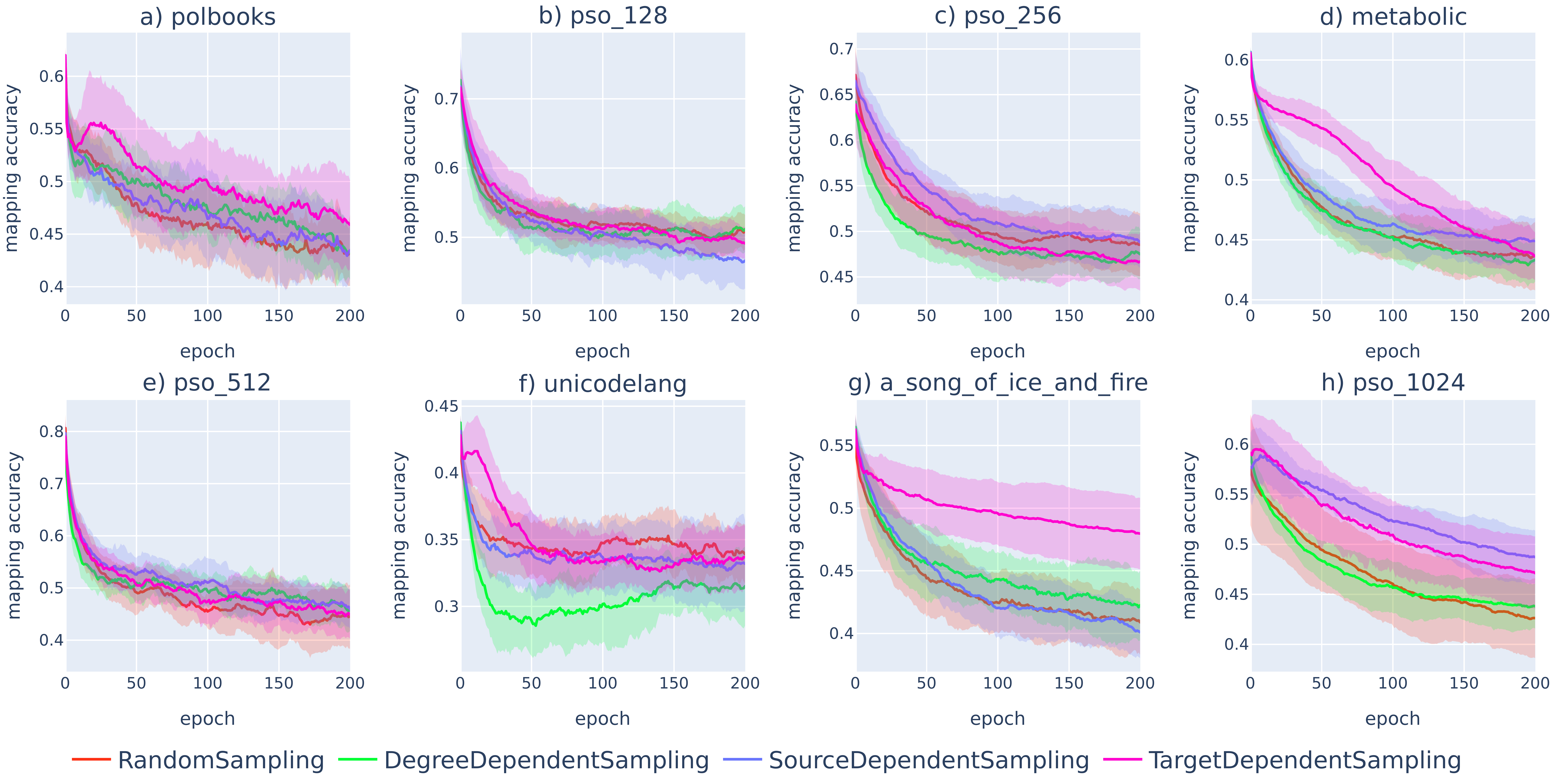}
    \caption{ {\bf The average mapping accuracy over 16 instances as a function of the number of epochs during the optimisation when the initial embedding was obtained with Mercator\cite{Mercator}.} The colour indicates the annealing scheme, the shaded region around the curves is showing the standard deviation and the network is indicated in the panel title.
    }
    \label{fig:MA_mercator}
\end{figure}
\begin{figure}[h!]
    \centering
    \includegraphics[width=\textwidth]{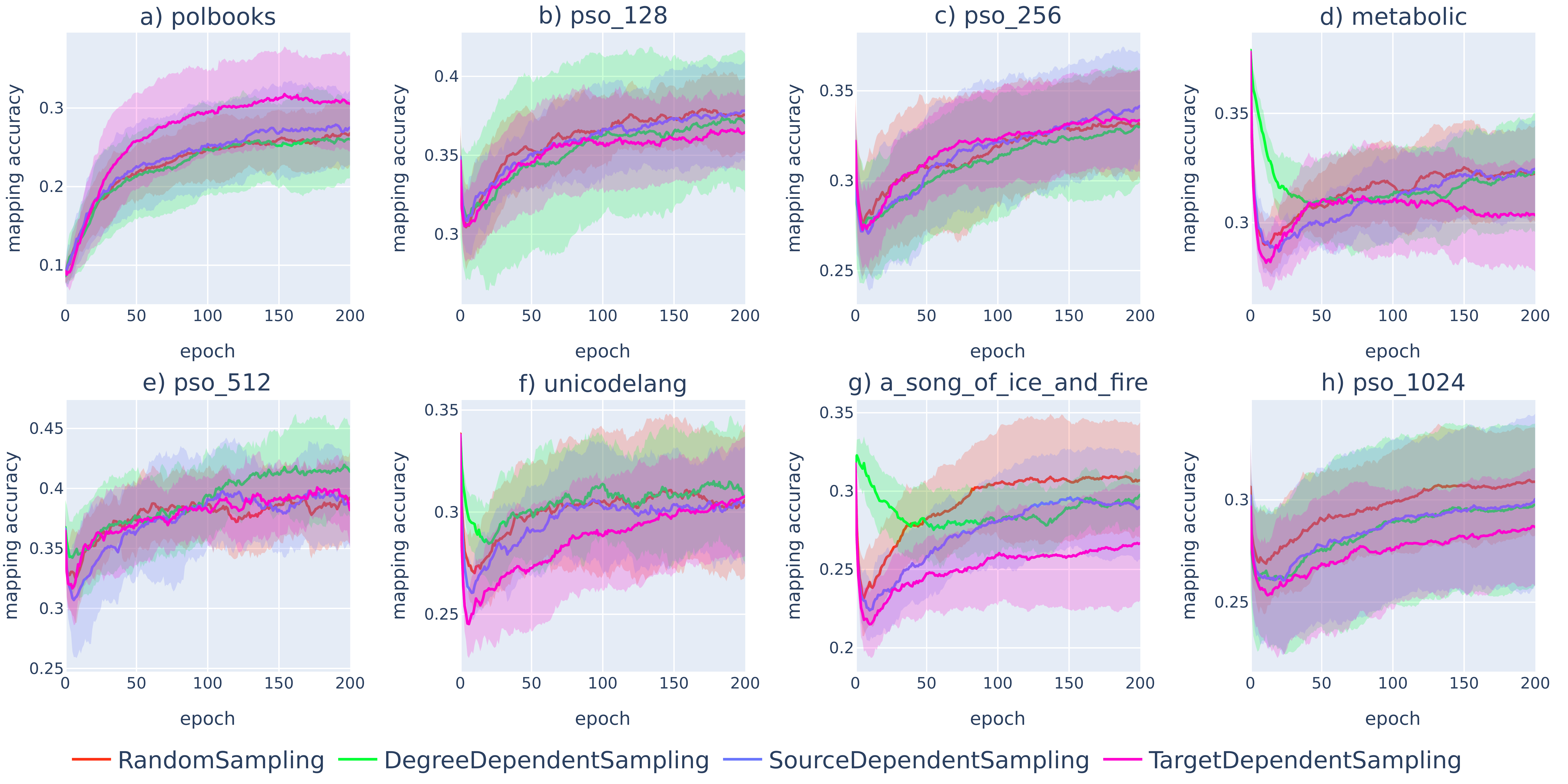}
    \caption{ {\bf The average mapping accuracy over 16 instances as a function of the number of epochs when the optimisation starts from a random embedding.} The colour indicates the annealing scheme, the shaded region around the curves is showing the standard deviation and the network is indicated in the panel title.
    }
    \label{fig:MA_random}
\end{figure}

In Fig.~\ref{fig:MA_mercator}.~ we show the mapping accuracy for the studied networks when starting the optimisation from node coordinates generated by Mercator\cite{Mercator}, whereas Fig.~\ref{fig:MA_random}. displays the same results for random initial coordinates. Since Mercator is usually producing embeddings with good quality measures and our optimisation is focusing on a different quality score, the mapping accuracy shows a decreasing tendency for all networks in Fig.~\ref{fig:MA_mercator}, and the magnitude of the change can be quite significant. Hence, the loss in the mapping accuracy can be viewed as a price we have to pay when optimising the a Mercator based embedding with respect to the success ratio. Nevertheless, according to Fig.~\ref{fig:MA_random}.~, the mapping accuracy tends to be either increasing, or remains more or less constant when we start the simulated annealing from a random initial embedding. 

\subsection{Quality scores related to graph reconstruction}

The graph reconstruction problem can be viewed as a task designed to examine to what extent is it possible to distinguish between connected and un-connected node pairs based on the geometric distance between their endpoints. Intuitively, we may imagine deleting all links from the embedded network, and then trying to reconstruct the graph by connecting $L$ number of node pairs based on the connection probability given by the hyperbolic coordinates (where $L$ is equal to the initially existing number of links). Since the connection probability is a monotonously decreasing function of the distance in hyperbolic networks, in practice this is equivalent to first ranking the node pairs according to their hyperbolic distance, and then connecting the $L$ node pairs that appear at the top of the list. 
\begin{figure}[h!]
    \centering
    \includegraphics[width=\textwidth]{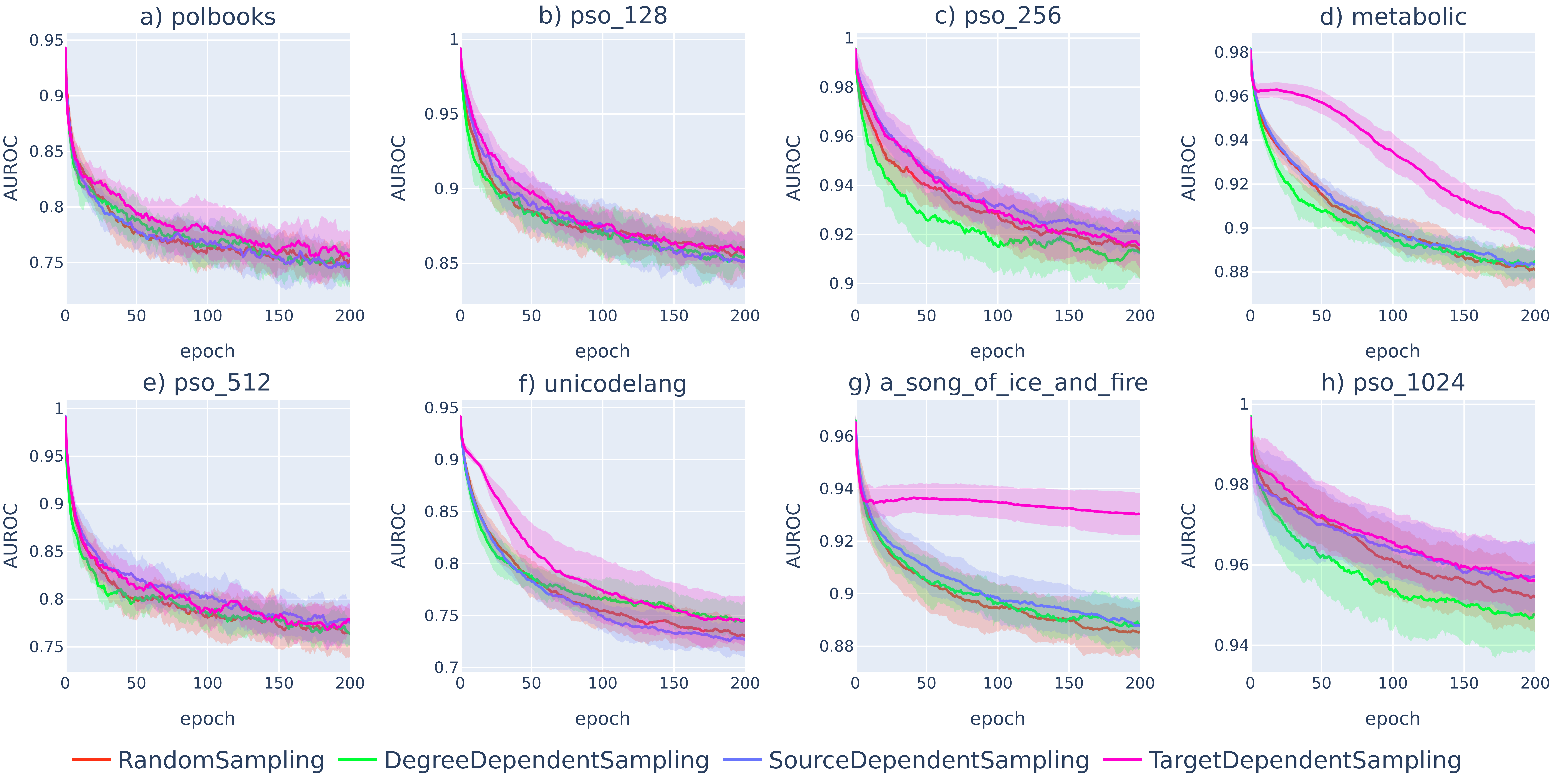}
    \caption{ {\bf The average AUROC  over 16 instances as a function of the number of epochs when the optimisation starts from an embedding obtained with Mercator\cite{Mercator}.}  The colour indicates the annealing scheme, the shaded region around the curves is showing the standard deviation and the network is indicated in the panel title. 
    }
    \label{fig:AUROC_mercator}
\end{figure}
\begin{figure}[h!]
    \centering
    \includegraphics[width=\textwidth]{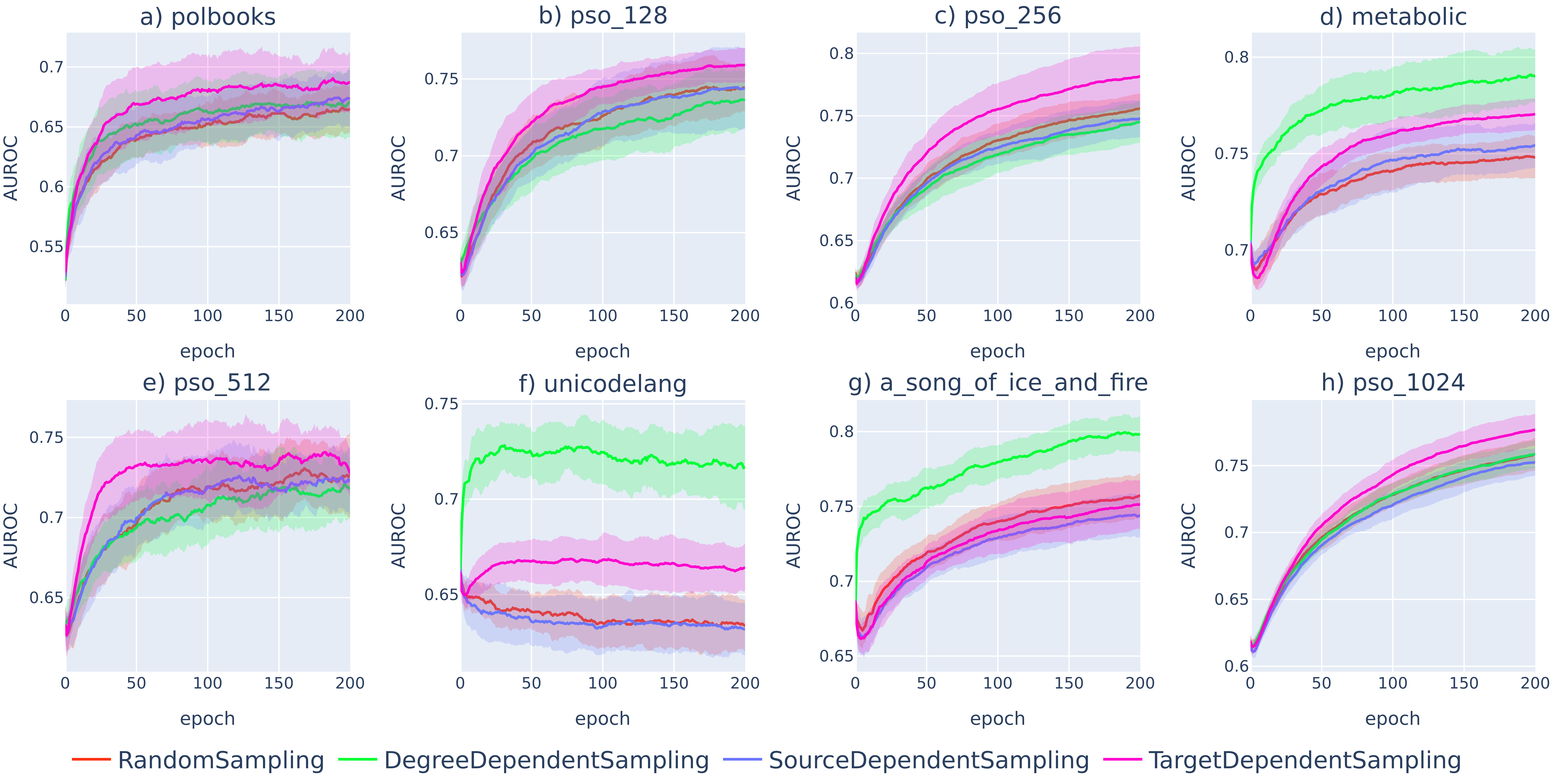}
    \caption{ {\bf The average AUROC over 16 instances as a function of the number of epochs when the optimisation starts from a random embedding.} The colour indicates the annealing scheme, the shaded region around the curves is showing the standard deviation and the network is indicated in the panel title.
    }
    \label{fig:AUROC_random}
\end{figure}

In a more general setting, the number of "reconstructed" links does not have to match the actually existing number of links in the system, and instead it can serve as a parameter that can be changed starting from $0$ up to the maximum possible number of links given by $N(N-1)/2$ in a network with $N$. In such a process, the receiver operating characteristic curve (ROC) is defined by plotting the ratio between the number of correctly reconstructed links and the number of existing links in the original network as a function of the ratio between the number of all reconstructed links (both correct and "false positive") and the maximum number of possible links\cite{AUROCmeaningInGeneral,AUROCmeaningInLinkPred}. The ROC always starts at $[0,0]$ and ends at $[1,1]$, and for random prediction takes the form of a simple straight line between these endpoints. To summarise the behaviour of the receiver operating characteristic curve it is a standard practice to take the area under the ROC, which is usually denoted by $\mathrm{AUROC}\in[0,1]$, as was done in previous studies of graph reconstruction and link prediction\cite{AUROCmeaningInLinkPred,descriptionOfMeasuresOfGraRecAndLinkPred}. (The AUROC value of a simple random predictor takes the value of AUROC=$1/2$).

In Fig.~\ref{fig:AUROC_mercator}.~ we show the AUROC values observed when optimising embeddings obtained with Mercator\cite{Mercator}, whereas Fig.~\ref{fig:AUROC_random}. displays the results when we start the annealing procedure from random embeddings. Similarly to the mapping accuracy in Fig.~\ref{fig:MA_mercator}., the AUROC shows a decreasing tendency for all networks in Fig.~\ref{fig:AUROC_mercator}. This again shows that Mercator\cite{Mercator} generates embeddings with high quality scores, and as our algorithm is optimising solely for the success ratio, the value of other quality indicators is likely to decrease when applying our framework on the output by Mercator. In contrast, the curves have almost always an increasing nature in Fig.~\ref{fig:AUROC_random}.~, indicating that compared to a fully random layout that is agnostic to the geometry, our simulated annealing procedure can achieve a better configuration even according to AUROC value, in spite that it is not designed to optimise this score.

A further quality of interest in graph reconstruction is given by the precision, defined as the fraction of the correctly predicted existing links among all reconstructed links\cite{descriptionOfMeasuresOfGraRecAndLinkPred}. Similarly to the ROC, we can also plot the precision as a function of the ratio between the number of restored links and the number of all possible node pairs, resulting in the so-called precision-recall curve. For a random predictor, this curve is corresponding to a constant curve at a value given by $\frac{2L}{N(N-1)}$, where $L$ is the number of links and $N$ is the number of nodes in the original network. The area under the precision-recall curve $\mathrm{AUPR}\in(0,1]$ provides a further important measure that can be used for quantifying the quality of a network embedding\cite{descriptionOfMeasuresOfGraRecAndLinkPred}.

\begin{figure}[h!]
    \centering
    \includegraphics[width=\textwidth]{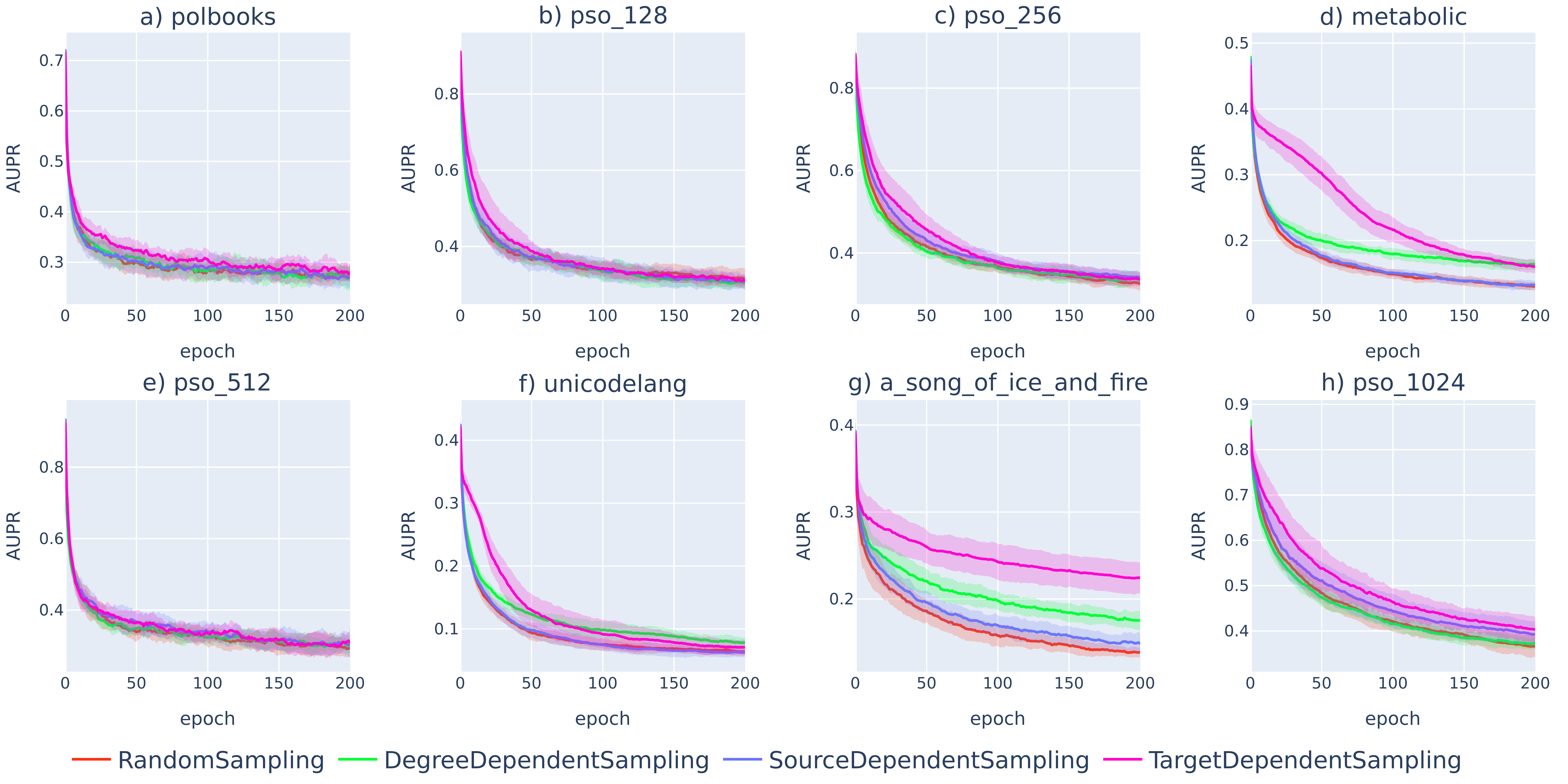}
    \caption{ {\bf The average AUPR over 16 instances as a function of the number of epochs when the optimisation starts from an embedding obtained with Mercator\cite{Mercator}.} The colour indicates the annealing scheme, the shaded region around the curves is showing the standard deviation and the network is indicated in the panel title. 
    }
    \label{fig:AUPR_mercator}
\end{figure}
\begin{figure}[h!]
    \centering
    \includegraphics[width=\textwidth]{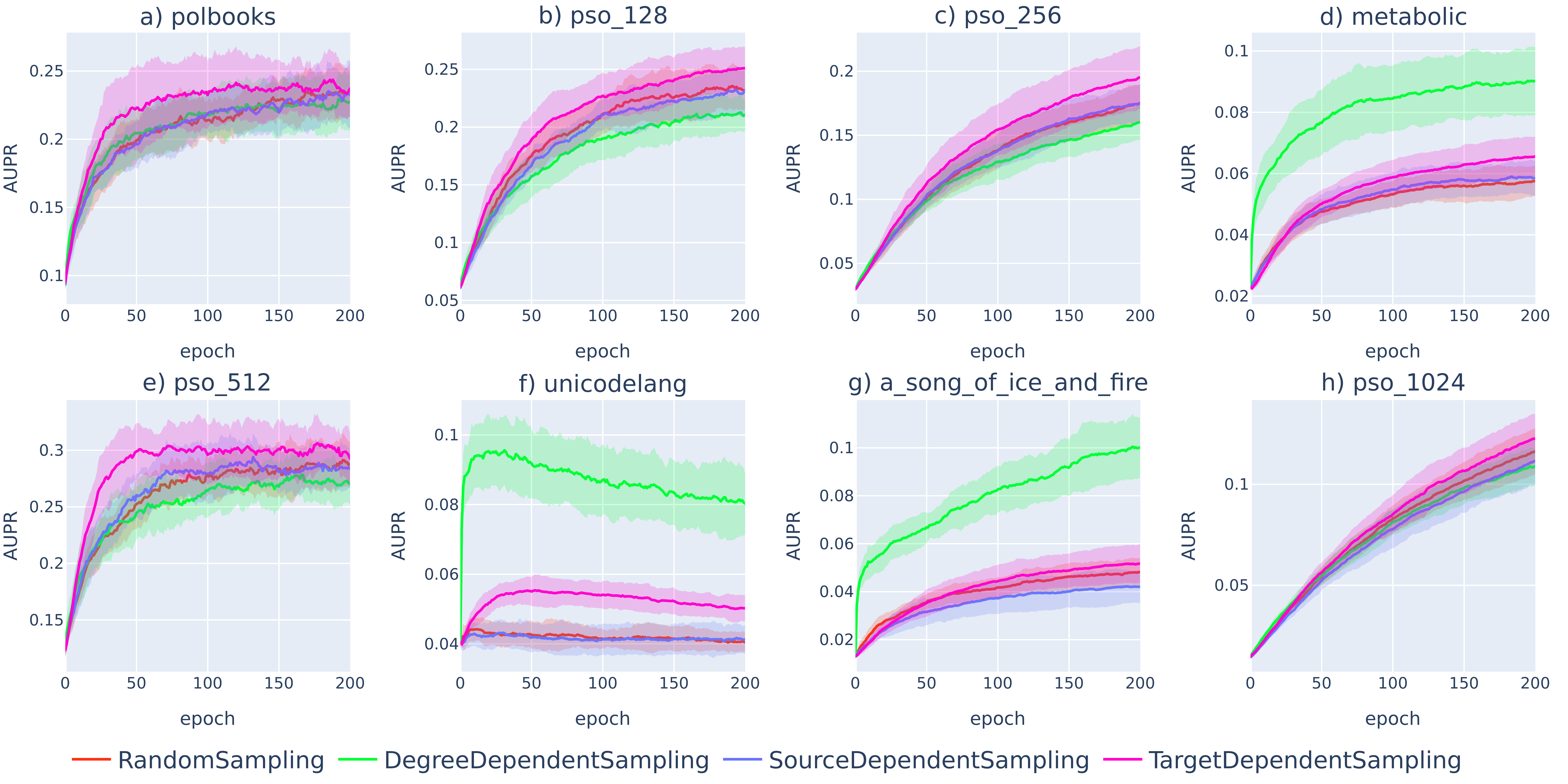}
    \caption{ {\bf The average AUPR over 16 instances as a function of the number of epochs when the optimisation starts from  a random embedding.} The colour indicates the annealing scheme, the shaded region around the curves is showing the standard deviation and the network is indicated in the panel title. 
    }
    \label{fig:AUPR_random}
\end{figure}

In Fig.~\ref{fig:AUPR_mercator}.~ we display the AUPR when starting the simulated annealing from embeddings obtained with Mercator\cite{Mercator}. Similarly to the AUROC score, we can observe a significant decrease in all of the cases, showing that our optimisation can have disadvantageous side affects when approaching the quality of an embedding from multiple perspectives.
The analogous results for the AUPR in optimisation experiments starting from random initial coordinates are presented in Fig.~\ref{fig:AUPR_random}. These plots show increasing AUPR curves, achieving a notable improvement in some of the cases. 
We note that similarly to the results related to the greedy congruence in Fig.\ref{fig:GC_random}., the different sampling methods may lead to different performance in terms of the AUROC and AUPR scores as well in some of the networks. 

\subsection{Quality scores related to greedy routing}

\begin{figure}[h!]
    \centering
    \includegraphics[width=\textwidth]{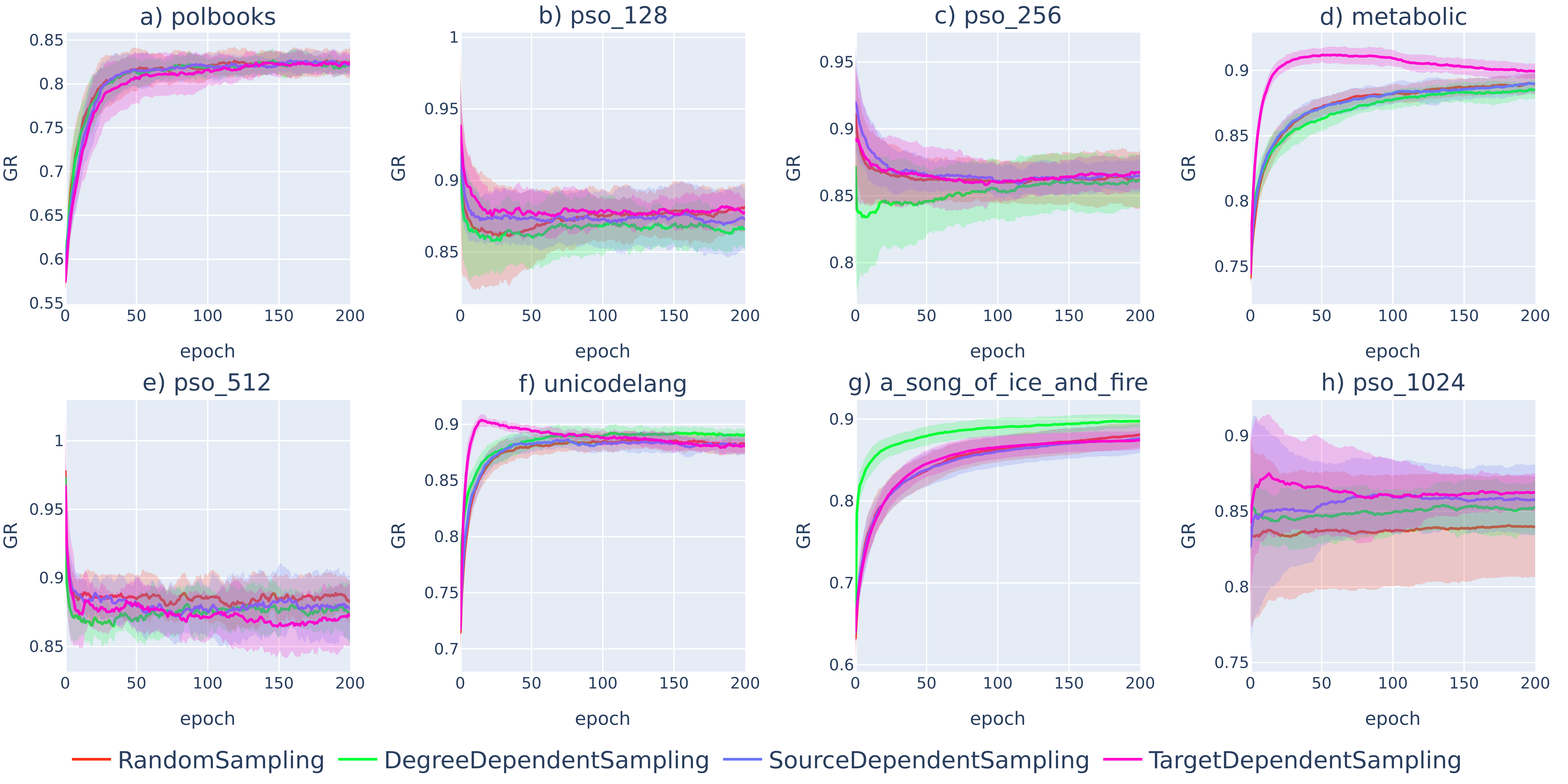}
    \caption{ {\bf The average of the greedy routing score, GR, over 16 instances as a function of the number of epochs when the optimisation starts from an embedding obtained with Mercator\cite{Mercator})}. The colour of the curves encode the annealing scheme, the shaded region around the curves is indicating the standard deviation and the network is indicated in the panel title. 
    }
    \label{fig:GR_Mercator}
\end{figure}

Besides the success ratio studied in the main paper, further quality measures can be defined for quantifying the greedy navigability of networks embedded in geometric spaces. An important example is given by the greedy routing score~\cite{coalescentEmbedding}, as defined in Eq.(1) in the main paper. Instead of focusing only on the fraction of the successful greedy paths, this measure incorporates information about the lengths of the paths as well. In Fig.~\ref{fig:GR_Mercator}. we show the behaviour of this score when the simulated annealing scheme starts from an embedding obtained with Mercator\cite{Mercator}. Interestingly, for the studied real networks the GR-score is improving during the optimisation, whereas in the case of the PSO networks it is either constant, or becomes constant after a small drop at the beginning of the simulated annealing process. 
\begin{figure}[h!]
    \centering
    \includegraphics[width=\textwidth]{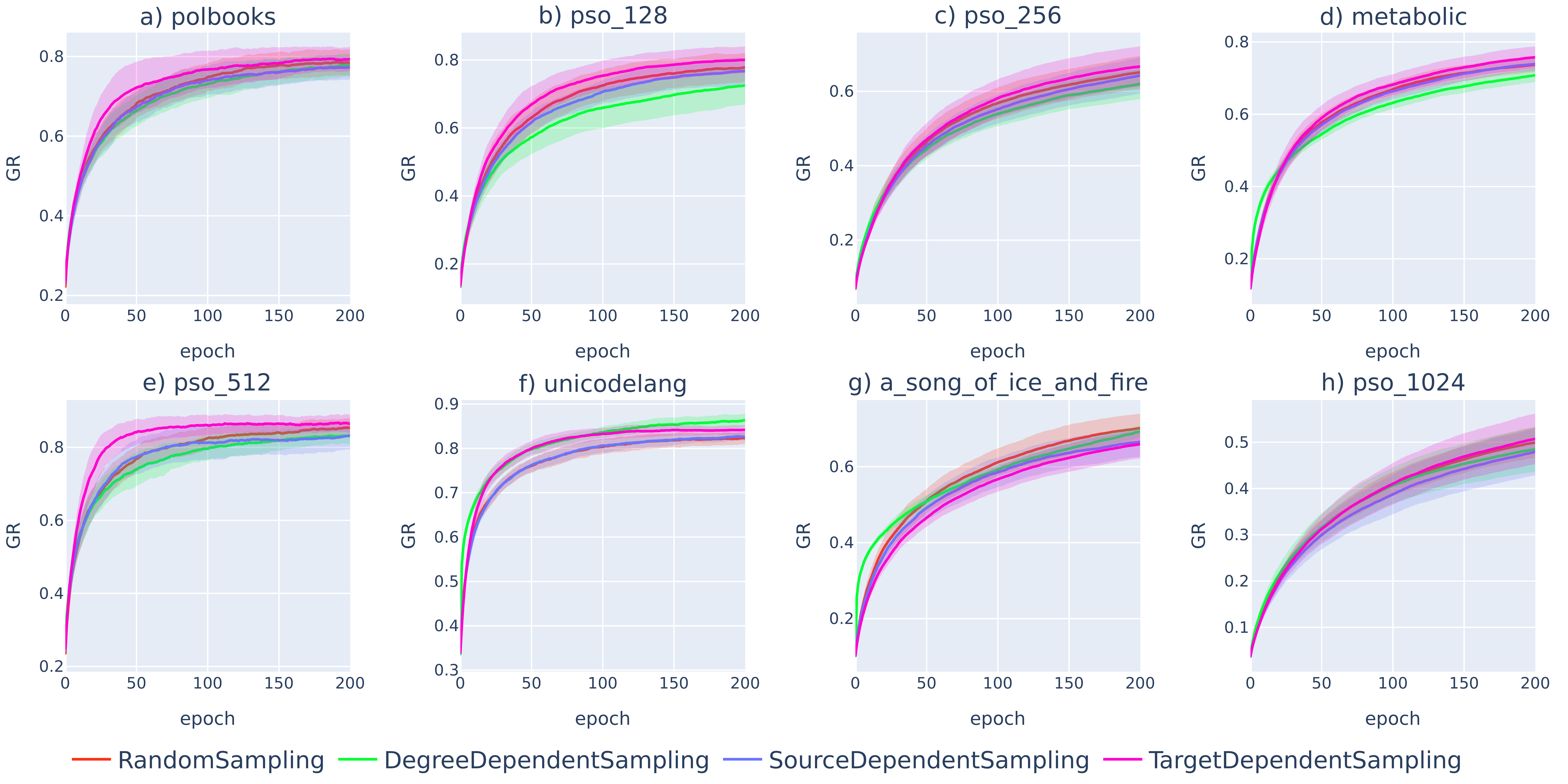}
    \caption{ {\bf The average GR over 16 instances as a function of the number of epochs when the optimisation starts from  a random embedding.} The colour indicates the annealing scheme, the shaded region around the curves is showing the standard deviation and the network is indicated in the panel title. 
    }
    \label{fig:GR_random}
\end{figure}

In parallel, Fig.~\ref{fig:GR_random}.~ depicts the GR-score when the optimisation is started from a random embedding, displaying a major improvement in the score for all networks during the simulated annealing process. The likely reason behind this is that the GR-score and the GS-score are closely related, and when optimising a random initial state with respect to the GS-score, most of the implemented displacements increase the GR-score as well.

In addition, a third quality score that was proposed to quantify the greedy navigability of a network is given by the greedy routing efficiency\cite{Carlo_Nat_coms_hyp_congruency}, comparing the geometric distances and the topological shortest paths. This measure can be formulated as
\begin{equation}
    GE(\{r_i, \theta_i\})
    =
    \frac{1}{N (N-1) - L}
    \sum_{i=1}^{N} \sum_{\substack{j=1 \\ j \neq i \\ j \notin N(i)}}^{N} \frac{DIST(i, j)}{PGRP(i, j)},
    \label{eq:GE_eff}
\end{equation}
where the summation runs over all nonadjacent pair of nodes, $DIST(i, j)$ denotes the geometric distance between $i$ and $j$, $PGRP(i, j)$ is corresponding to the length of the projected greedy routing path between the same pair of nodes, $N$ gives the total number of nodes and $L$ stands for the number links in the network.

\begin{figure}[h!]
    \centering
    \includegraphics[width=\textwidth]{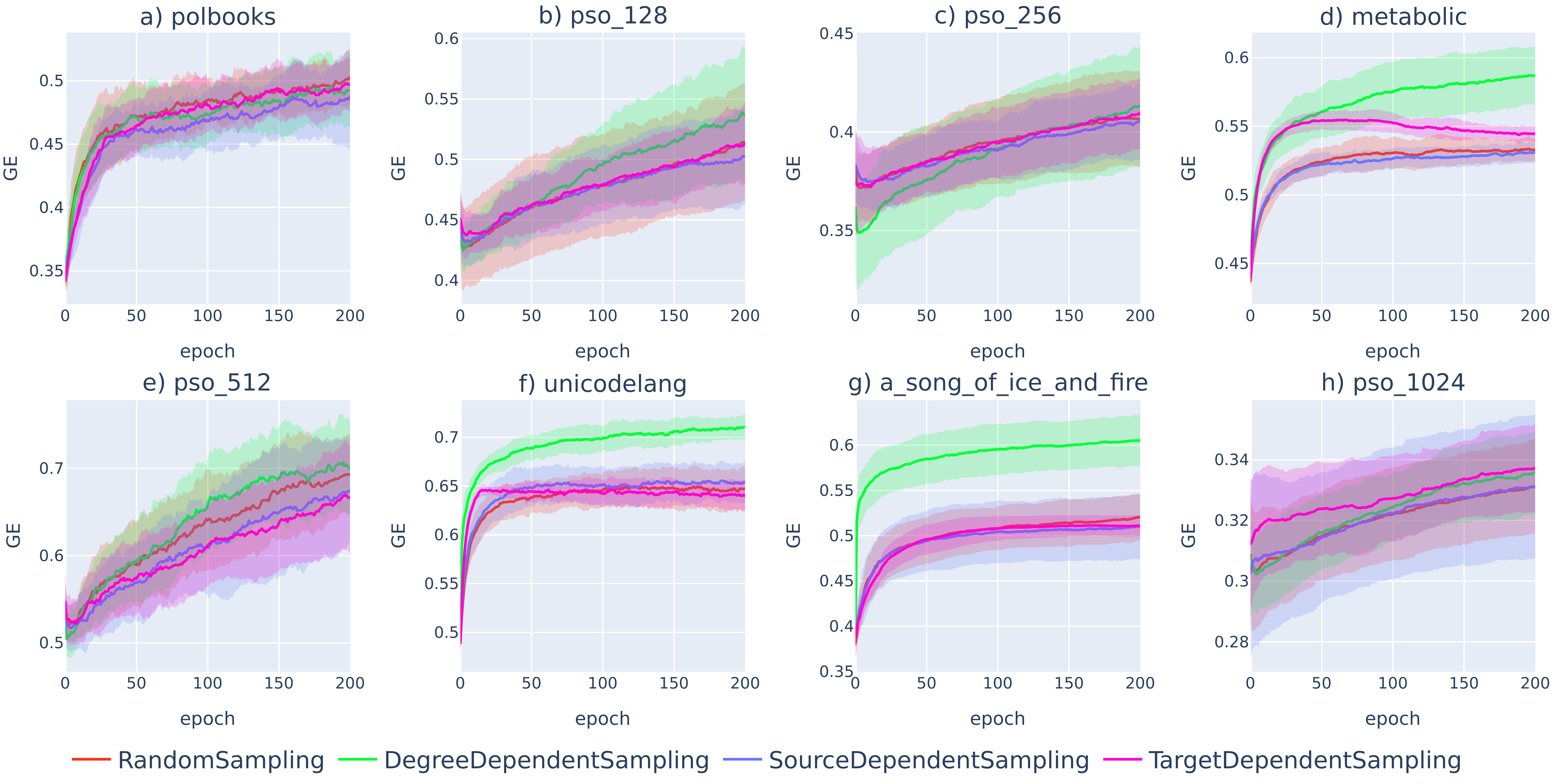}
    \caption{ {\bf The average of GE over 16 instances as a function of the number of epochs when the optimisation starts from an embedding obtained with Mercator\cite{Mercator})}. The colour of the curves encode the annealing scheme, the shaded region around the curves is indicating the standard deviation and the network is indicated in the panel title. 
    }
    \label{fig:GE_Mercator}
\end{figure}
\begin{figure}[h!]
    \centering
    \includegraphics[width=\textwidth]{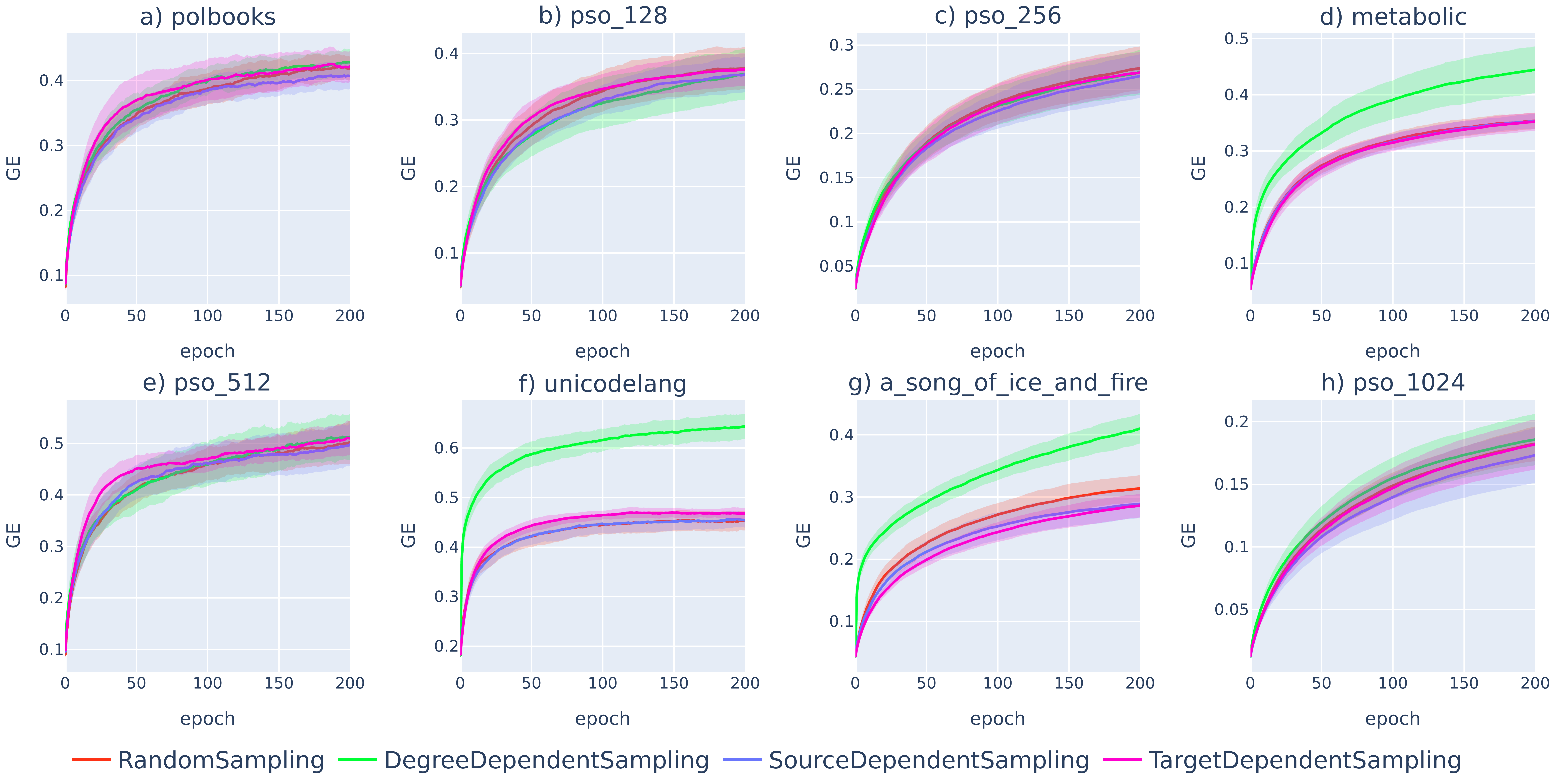}
    \caption{ {\bf The average GE over 16 instances as a function of the number of epochs when the optimisation starts from  a random embedding.} The colour indicates the annealing scheme, the shaded region around the curves is showing the standard deviation and the network is indicated in the panel title. 
    }
    \label{fig:GE_random}
\end{figure}

In Fig.~\ref{fig:GE_Mercator}.~ we present the results for GE in simulated annealing experiments starting from embeddings obtained with Mercator\cite{Mercator}, whereas Fig.~\ref{fig:GE_random}.~ depicts the same results obtained when we start from random initial embeddings. The tendency of the curves is showing an increasing tendency in all experiments, and a considerable improvement can be observed in most of the cases over the iterations. This means that our optimisation framework is likely to find spatial configurations that are more advantageous not only from the point of view of the success of the greedy paths, but also in terms of the efficiency.

Finally, when comparing the performance results for the different sampling methods in terms of the scores related to greedy routing, again, clear separation of the curves can be observed for some of the networks. By taking together Figs.~\ref{fig:GC_random}.,~\ref{fig:AUROC_random}.,~\ref{fig:AUPR_random}.,~\ref{fig:GE_Mercator}. and Fig.~\ref{fig:GE_random}., it seems that the degree based sampling can achieve somewhat better results according to some of the quality scores for the metabolic, unicodelang and fictional character networks. However, since this effect is not seen for the other networks, and not for all quality scores in the case of these 3 networks, it is likely to be the result of a particular interplay between the specific structure of these networks, the annealing framework and the nature of the given quality measure. 

\clearpage

\bibliographystyle{unsrt}

\bibliography{gr_paper}

\end{document}